\documentclass[12pt,twoside]{article}

\pdfoutput=1

\usepackage[usenames]{color}
\usepackage{lscape}
\usepackage{amssymb}
\usepackage{rotating}
\usepackage[T1]{fontenc}
\usepackage{graphicx}
\usepackage{fancyhdr}
\usepackage{amsmath}
\usepackage{ae}
\usepackage{aecompl}
\usepackage{hyperref}

\def\nn{\nonumber \\}

\def\hgamgam{h\to \gamma\gamma}

\newcommand{\GeV}{{\rm GeV}}

\newlength{\dinwidth}
\newlength{\dinmargin}
\begin{document}

\thispagestyle{empty}

\vspace*{1cm}

\centerline{\Large\bf Light staus and enhanced Higgs diphoton rate } 
\vspace*{2mm}
\centerline{\Large\bf with non-universal gaugino masses} 
\vspace*{2mm}
\centerline{\Large\bf and SO(10) Yukawa unification} 

\vspace*{5mm}

\vspace*{5mm} \noindent
\vskip 0.5cm
\centerline{\bf
Marcin Badziak\footnote[1]{mbadziak@fuw.edu.pl},
Marek Olechowski\footnote[2]{Marek.Olechowski@fuw.edu.pl},
Stefan Pokorski\footnote[3]{Stefan.Pokorski@fuw.edu.pl}
}
\vskip 5mm

\centerline{\em Institute of Theoretical Physics, Faculty of Physics,
University of Warsaw} 
\centerline{\em ul.\ Ho\.za 69, PL--00--681 Warsaw, Poland}

\vskip 1cm

\centerline{\bf Abstract}
\vskip 3mm

It is shown that substantially enhanced Higgs to diphoton rate induced by 
light staus with large left-right mixing in MSSM requires at the GUT scale 
non-universal gaugino masses with bino and/or wino lighter than gluino. The 
possibility of such enhancement is investigated in MSSM models with arbitrary 
gaugino masses at the GUT scale with additional restriction of top-bottom-tau 
Yukawa unification, as predicted by minimal SO(10) GUTs. Many patterns of 
gaugino masses leading to enhanced Higgs to diphoton rate and the Yukawa 
unification are identified. Some of these patterns can be accommodated in a 
well-motivated scenarios such as 
mirage mediation or
SUSY breaking $F$-terms  
being a non-singlet of SO(10). Phenomenological implications of a scenario 
with non-universal gaugino masses generated by a mixture of the singlet 
$F$-term and the $F$-term in a 24-dimensional representation of SU(5) 
$\subset$ SO(10) are studied in detail.  
Possible non-universalities of other soft terms generated by 
such F-terms are discussed. 
The enhancement of Higgs to diphoton 
rate up to 30\% can be obtained in agreement with all phenomenological 
constraints, including vacuum metastability bounds. The lightest sbottom 
and pseudoscalar Higgs are within easy reach of the 14 TeV LHC. 
The LSP can be either bino-like or wino-like. The thermal relic abundance 
in the former case may be in agreement with the cosmological data 
thanks to efficient stau coannihilation.

\newpage

\section{Introduction}

Discovery of a Higgs boson with mass around 125 GeV is now firmly established 
\cite{Atlas_discovery,CMS_discovery}. On the other hand, it remains unclear 
whether the discovered boson is the Standard Model (SM) Higgs boson. Even 
though the measured values of the Higgs signal, $\mu_i$, in most decay channels 
are within $1\sigma$ from the SM prediction, the errors are still rather 
large, of about 20--30\%, even in the best-measured channels such as 
$\gamma\gamma$, $WW^*$ and $ZZ^*$ 
\cite{Atlas_gamma,Atlas_WWZZ,CMS_gamma,CMS_WWZZ}. Moreover, there are some 
anomalies in the LHC data. Particularly interesting is an excess of events in 
the $\gamma\gamma$ channel observed by ATLAS. The fitted number of signal events in this channel is $2\sigma$ above the SM 
prediction \cite{Atlas_gamma}. 
Similar excess was observed before also in the CMS data \cite{CMS_gamma_old} 
but it disappeared after analysing all data collected in the 7 TeV and 8 TeV 
LHC runs \cite{CMS_gamma}.

In general, there are many ways to enhance the Higgs signal strength in the 
$\gamma\gamma$ channel. One possibility is to have the Higgs coupling to 
$b\bar{b}$ smaller than in the SM because this leads to a reduction of the 
total Higgs decay width and, as a result, increases the Higgs branching ratios 
into other states. Since the $b\bar{b}$ channel dominates decays of the 125 
GeV Higgs in the SM even small decrease of the $hb\bar{b}$ coupling gives 
non-negligible enhancement in other channels. Such effect is possible, for 
example, in the minimal supersymmetric standard model (MSSM) due 
to mixing between the the two CP-even Higgs bosons 
\cite{DjouadiMSSMreview}.\footnote{
  In the MSSM the Higgs coupling to $b\bar{b}$ can be reduced only if loop 
  corrections to off-diagonal entries of the Higgs mass matrix from 
  the-third-generation sfermion sector are significant \cite{Carena_stau}. 
  On the other hand, in the NMSSM this effect can be present already 
  at the tree level because of the mixing of the two Higgs doublets 
  with the singlet 
  \cite{Ellwanger_reducedhbb,Cao}.} 
However, this effect enhances the Higgs signal strength in the $WW^*$ and
$ZZ^*$ by the same amount as in the $\gamma\gamma$ channel unless partial
Higgs decay widths are non-universally modified. There are no hints in the LHC
data for any correlation between the Higgs signal strength in the $WW^*/ZZ^*$
channels and the $\gamma\gamma$ channel so it seems more likely that the
enhancement in the $\gamma\gamma$ channel is due to enhanced $\Gamma(h\to
\gamma\gamma)$. That is why in this paper we study the possibility that the Higgs
signal strength is enhanced in the $\gamma\gamma$ channel while other channels
are SM-like.

Since in the SM the Higgs decays into $\gamma\gamma$ only at loop level, 
substantial corrections to $\Gamma(h\to \gamma\gamma)$ are possible due to 
new electromagnetically charged states with sizeable couplings to the Higgs 
\cite{Carena_diphoton}. 
Many models have recently appeared in the literature in which the $\hgamgam$ rate is enhanced due to new charged scalars, gauge bosons or vector-like fermions. For representative examples of such scenarios see e.g.~Refs.~\cite{diphoton_scalar,diphoton_boson,diphoton_fermion}. 
In this paper we focus on the possibility that 
such new states are supersymmetric and study enhanced $\hgamgam$ rate 
in the MSSM. Such possibility is very limited in the MSSM since only the 
third-generation sfermions \cite{Carena_stau} and charginos 
\cite{gamma_chargino} may couple to the Higgs strongly enough to have 
non-negligible impact on $\Gamma(h\to\gamma\gamma)$. The most attractive 
possibility is that the $\hgamgam$ rate is enhanced by light staus 
with large left-right mixing
\cite{Carena_stau}.\footnote{
  The impact of light staus on the $\hgamgam$ rate
  was also investigated in extensions of the MSSM \cite{Staub_stau}.
}

Effects of staus on $\hgamgam$ rate have been studied so far from a low-energy 
perspective. The purpose of the present work is to show that SUSY spectrum 
with light staus enhancing $\hgamgam$ rate may emerge from a well-motivated 
high-energy scenario. In particular, we point out that GUT-scale boundary 
conditions for the MSSM soft terms that may lead to enhanced $\hgamgam$ rate 
have to include non-universal gaugino masses with bino and/or wino lighter 
than gluino.

Strong left-right mixing in the stau sector, as required by enhanced 
$\hgamgam$ rate,  strongly prefers models with large $\tan\beta$. Large values 
of $\tan\beta$ can explain the observed hierarchy between the top and bottom 
masses \cite{mosp_tbuni} and are predicted in minimal SO(10) models since they 
impose unification of top, bottom and tau Yukawa couplings at the GUT scale 
\cite{YU_tbtau,bop_F24,mbks}. Therefore, these models are perfect candidates 
to accommodate enhanced $\hgamgam$ rate. Unification of the Yukawa couplings 
by itself is very sensitive to low-energy SUSY threshold corrections, mainly because large values of $\tan\beta$ induce big threshold correction to the bottom mass 
\cite{Hall,Carena}, so it gives constraints for soft SUSY breaking terms at the 
GUT scale, as well as for low-energy SUSY spectrum.\footnote{The gauge coupling unification also has some sensitivity to the MSSM spectrum but as long as gaugino and higgsino masses are not far above ${\mathcal O}(10\ {\rm TeV})$ the gauge couplings unify within a few percent, see e.g. Ref.~\cite{gaugeuni}. On the other hand, for large $\tan\beta$ the correction to the bottom mass varies between a few and several tens of percent depending on some hierarchies in the SUSY spectrum.} 
In particular, it has been 
known for a long time that Yukawa unification requires to go beyond universal 
soft SUSY breaking terms at the GUT scale \cite{Olechowski}. The recent 
measurement of the Higgs mass of about 125 GeV gives additional constraints on 
the SUSY spectrum \cite{YU_125Higgs}. Implications of the 125 GeV Higgs 
for SO(10) Yukawa unification were recently reviewed in Ref.~\cite{YUreview}. 
The most important effect of the measured Higgs mass is that most of 
the SUSY spectrum is pushed above 1 TeV, and typically sparticle masses are in a few TeV range. Given the above constraints 
it is not obvious from the start whether sufficiently light and 
strongly-mixed staus leading to enhanced $\hgamgam$ rate are possible 
in SO(10) models. In this paper we show that enhanced $\hgamgam$ rate 
can be obtained in agreement with top-bottom-tau Yukawa unification.  
Assuming $D$-term splitting of scalar masses, which generically arises in
spontaneously broken SO(10) models \cite{Dterm} and is needed to account for
proper radiative electroweak symmetry breaking (REWSB) \cite{Murayama}, we
identify patterns of gaugino masses that may allow for enhanced $\hgamgam$
rate. We point out that appropriate patterns of gaugino masses can be
accommodated in mirage mediation which often appears in low-energy limits of
string theories \cite{mirage1}-\cite{mirage7}. Enhanced $\hgamgam$ rate is
possible also if SUSY breaking $F$-term that contributes to gaugino masses is
a combination of a singlet and a non-singlet representation of SO(10). We
investigate the latter scenario in more detail focusing on the most 
attractive case with the gaugino masses generated by a mixture of the 
singlet $F$-term and the $F$-term in ${\bf 24}\subset{\bf 54}$ of 
$SU(5)\subset SO(10)$.

The rest of the paper is organized as follows. In Section \ref{sec:Rgam} 
we review the Higgs to diphoton decays with a special emphasis on 
effects from light staus and argue that at the GUT scale the bino or wino 
mass should be smaller than the gluino mass in order to substantially 
enhance $\hgamgam$ rate. In Section \ref{sec:YU} 
we review necessary conditions for top-bottom-tau Yukawa unification and 
identify patterns of gaugino masses that can accommodate enhanced $\hgamgam$ 
rate in this class of models. In Section \ref{sec:F1F24} we investigate 
in detail a model with the gaugino masses generated by the singlet $F$-term 
and the $F$-term in ${\bf 24}\subset{\bf 54}$ of $SU(5)\subset SO(10)$. 
First we carefully discuss possible patterns of other soft terms - 
scalar masses and trilinear terms. Later we concentrate on the 
predictions of the model for the MSSM spectrum. 
Finally, we summarize our findings in Section \ref{sec:concl}.

\section{Enhanced Higgs to diphoton rate in MSSM}
\label{sec:Rgam}

In the SM, $\hgamgam$ is a loop mediated process with a dominant contribution
from $W^{\pm}$ and a smaller, but non-negligible, contribution from the top
quark \cite{DjouadiSMreview}: 
\begin{equation}
   \Gamma^{\rm SM}(\hgamgam) 
\approx 
\frac{G_\mu\alpha^2 m_{h}^3}{128\sqrt{2}\pi^3}  
\bigg| A_1 \left(\frac{m_h^2}{4m_W^2}\right) 
+ \frac{4}{3} A_{1/2}\left(\frac{m_h^2}{4m_t^2}\right) \bigg|^2 \,,
\end{equation}
where the amplitudes for the spin-1 and spin-1/2 particle contributions are 
given at the leading order by: 
\begin{align}
\label{A12}
 A_{1/2}(x)  &=  2 [x +(x-1)\arcsin^2 \sqrt{x}]\, x^{-2} \,, \\  
\label{A1}
 A_1(x)      &=  - [2x^2 +3x+3(2x -1)\arcsin^2 \sqrt{x}]\, x^{-2} \,,
\end{align}
with $x_i=m^2_h/4m^2_i$ and $m_i$ denoting the mass of the particle running
in the loop.
In the MSSM, non-negligible contributions can originate from very light 
sfermions provided that their Yukawa couplings are large 
\cite{DjouadiMSSMreview}. Thus, the candidates for the modification 
of the Higgs diphoton rate are stops, and at large $\tan\beta$ also 
sbottoms and staus. Light stops and sbottoms with large left-right 
mixing can enhance $\Gamma(\hgamgam)$ but at the cost of even larger 
reduction of the Higgs production cross-section via gluon fusion
\cite{Djouadi_reducedcross}.\footnote{
  In principle, simultaneous enhancement of $\Gamma(\hgamgam)$ and the 
  production rate is possible for very light stops with left-right mixing 
  large enough to make the stop contribution to the amplitude of the gluon 
  fusion process more than two times larger than the corresponding 
  contribution from the top quark. However, for stop masses consistent 
  with the experimental constraints the required stop mixing gives negative 
  contribution to the Higgs mass so obtaining 125 GeV for this mass would be 
  very problematic.} 
On the other hand, light staus with large left-right mixing can enhance 
$\Gamma(\hgamgam)$ without affecting the Higgs production rate 
\cite{Carena_stau}. In the following Subsection we describe the effects of 
light staus on the Higgs decay rate to two photons in some more detail.

\subsection{Effects of light staus}
\label{subsec:staus}

The possibility of the $\hgamgam$ rate enhancement by light staus was first emphasized in Ref.~\cite{Carena_stau}. In order to better understand the origin of this effect we collect below the most relevant formulae.

For the 125 GeV Higgs, the MSSM Higgs diphoton rate (normalized to its SM 
value) including the stau effects is given by:  
\begin{equation}
 \frac{\Gamma(\hgamgam)}{\Gamma^{\rm SM}(\hgamgam)} 
\approx 
 \bigg| 1.28 c_V - 0.28 c_t - 0.15 c_{\gamma}^{\rm stau} \bigg|^2 \,,
\end{equation}
where $c_V$ and $c_t$ are the Higgs couplings to W boson and to top quark
(normalized to their SM values), respectively, and the coefficients 
in front of them were obtained using $A_i$ functions defined in 
eqs.~(\ref{A12})-(\ref{A1}). The contribution from staus is given 
by\cite{Carena_diphoton}:  
\begin{equation}
 c_{\gamma}^{\rm stau} 
=
 \sum_{i=1,2} 
g_{h \tilde{\tau}_i\tilde{\tau}_i}\frac{M_Z^2}{m_{\tilde{\tau}_i}^2 }
A_0\left(\frac{m_h^2}{4m_{\tilde{\tau}_i}^2}\right)
\end{equation}
with \footnote{
Different formulae for the Higgs-stau-stau couplings appear 
in the literature. The couplings we use in this paper agree, 
after taking into account the opposite sign convention  
for the stau mixing angle and an apparent misprint, with those 
from Ref.\ \cite{DjouadiMSSMreview}. They agree also with 
general results given in \cite{Carena_diphoton}. 
However, the formulae 
used in e.g.\ \cite{Rgam_g2} and \cite{AkulaNath} 
differ from ours (and one from the other) even if rewritten 
using the same sign convention.
}
\begin{align}
\label{htautau1}
 g_{h \tilde{\tau}_1\tilde{\tau}_1} 
&=
\cos 2\beta \left( -\frac{1}{2} \cos^2\theta_{\tilde\tau} 
+ \sin^2\theta_{\rm W}\cos 2\theta_{\tilde\tau} \right) 
+ \frac{m^2_\tau}{M^2_Z} 
- \frac{m_\tau X_{\tau}}{2 M_Z^2} \sin 2\theta_{\tilde \tau}\,,
\\
\label{htautau2}
g_{h \tilde{\tau}_2\tilde{\tau}_2} 
&=
\cos 2\beta \left( -\frac{1}{2} \cos^2\theta_{\tilde\tau} 
- \sin^2\theta_{\rm W}\cos 2\theta_{\tilde\tau} \right) 
+ \frac{m^2_\tau}{M^2_Z} 
+ \frac{m_\tau X_{\tau}}{2 M_Z^2} \sin 2\theta_{\tilde \tau}\,,
\end{align}
where
\begin{equation}
X_{\tau} = A_{\tau} - \mu\tan\beta
\end{equation}
and the form factor is given (for $x<1$ i.e. $m_h<2m_{\tilde{\tau}_i}$) by \cite{DjouadiMSSMreview}
\begin{equation}
 A_0(x)    =  - [x -\arcsin^2 \sqrt{x}]\, x^{-2} \,.
\end{equation}
The stau mixing angle, $\theta_{\tilde\tau}$, can be determined in 
terms of the soft, $m_L$ and $m_E$, 
and physical, $m_{{\tilde\tau}_{1}}$ and  $m_{{\tilde\tau}_{2}}$, 
stau masses using the following relations: 
\begin{equation}
 \cos 2\theta_{\tilde \tau} =
\frac{m^2_L - m^2_E}{m^{2}_{{\tilde\tau}_{1}}-m^{2}_{{\tilde\tau}_{2}}}
\,,\qquad
\sin 2\theta_{\tilde \tau} =
-\frac{2m_\tau X_{\tau}}{m^{2}_{{\tilde\tau}_{1}}-m^{2}_{{\tilde\tau}_{2}}}\,.
\end{equation}
We also note that the splitting between the stau masses equals  
\begin{equation}
 m_{\tilde{\tau}_2}^2 - m_{\tilde{\tau}_1}^2 
= \sqrt{(m_L^2-m_E^2)^2 +(2m_{\tau} X_{\tau})^2 }
\end{equation}
and is the smallest for $m^2_L=m^2_E$.

It is clear that the largest enhancement can be obtained for the smallest 
possible stau masses. Lower limits on the lightest stau vary between 
82 and 94 GeV depending on the stau mixing angle and on the mass of the 
LSP (if it is not strongly degenerate with stau) 
\cite{LEPstau}.\footnote{
  Those limits can be substantially weaker if the mass 
  splitting between the lighter stau and the LSP is below a few GeV.}
After taking into account these LEP constraints, the modification of 
$\Gamma(\hgamgam)$ coming from the first two terms in each of 
eqs.~(\ref{htautau1}) and (\ref{htautau2}) is at most $\mathcal{O}(5\%)$.  
Substantial enhancement is possible only if the last terms in these 
equations are large which requires that the staus are strongly mixed. 
It is useful to note that the modification of the Higgs decay rate to two photons is well approximated by:
\begin{equation}
\label{gamrate_app}
 \frac{\Gamma(\hgamgam)}{\Gamma^{\rm SM}(\hgamgam)}
\approx 
 \left|1+0.15 A_0\left(\frac{m_h^2}{4m_{\tilde{\tau}_1}^2}\right) 
\frac{m_{\tau}^2 X_{\tau}^2}{m_{\tilde{\tau}_1}^2 m_{\tilde{\tau}_2}^2} \right|^2
\end{equation}
where we assume that the Higgs couplings to W bosons and top quark are the 
same as in the SM, which is a very good approximation in the decoupling limit, 
$M_A\gg M_Z$. In the limit $x\to 0$ the function $A_0(x)\to 1/3$ but for a 
very light stau it gives additional enhancement e.g.\ 
$A_0(\frac{m_h^2}{4m_{\tilde{\tau}_1}^2}) \approx 1.3/3$ for 
$m_{\tilde{\tau}_1}=100\,$GeV. Eq.~(\ref{gamrate_app}) clearly demonstrates 
that a significant $\gamma\gamma$ rate enhancement is possible if both staus 
are very light and the value of $X_{\tau}$ is very large in order 
to compensate the suppression by the tau mass. For instance, the 
enhancement by 20\% requires $X_{\tau}\approx 70\, m_{\tilde{\tau}_2}$ 
for $m_{\tilde{\tau}_1}=100\,$GeV. This implies $|X_{\tau}|\gtrsim20\,$TeV. 
Therefore, large values of $\mu$ and $\tan\beta$ are necessary to 
obtain a substantial enhancement of the $\hgamgam$ rate.

The enhancement of $\Gamma(\hgamgam)$ cannot be, however, arbitrarily 
large because for too large values of $\mu\tan\beta$ the electroweak 
vacuum becomes metastable 
\cite{stability_Rattazzi,stability_Hisano,stability_Carena,stability_Kitahara}.
The range of $\mu\tan\beta$ for which the vacuum is stable (or metastable 
with the life-time longer than the age of the Universe) can be estimated 
using the following phenomenological formula \cite{stability_Kitahara}:
\begin{align}
\label{stability_cond}
   |\mu\tan\beta| < 56.9 \sqrt{m_{L} m_{E}} 
+ 57.1 \left(m_{L}+1.03 m_{E} \right)   
- 1.28 \times 10^4 \GeV \nn
+\frac{1.67 \times 10^6 \GeV ^2 }{m_{L}+m_{E} }  
- 6.41 \times 10^7 \GeV ^3 \left ( \frac{1}{m_{L}^2  } 
+ \frac{0.983}{m_{E}^2}  \right) .
\end{align}
which gives good approximation for $m_{L}$, 
$m_{E} \leq 2$ TeV.\footnote{Similar formula, valid only for 
smaller values of the soft stau masses, was given earlier in 
Ref.~\cite{stability_Hisano}.}  It was shown in Ref.~\cite{stability_Carena} 
that $\Gamma(\hgamgam)$ can be enhanced by up to 50\%  without 
violating the metastability bound.

\subsection{Enhanced {\boldmath $h\to\gamma\gamma$} rate from GUT-scale perspective}

Even though strongly-mixed light staus can lead to substantial enhancement 
of the $\hgamgam$ rate it has not been demonstrated so far that such 
pattern of SUSY spectrum can be obtained from a well-motivated high-energy 
model. In the following we argue that, under reasonable assumptions, 
the $\hgamgam$ rate cannot be substantially enhanced, say by 20\% or more, 
if gaugino masses are universal at the GUT scale.

Since enhancing the $\gamma\gamma$ rate requires the lightest stau mass to 
be around 100 GeV, avoiding a charged LSP implies $|\mu|$, $|M_1|$ or $|M_2|$ 
at the EW scale to be at most $\sim\mathcal{O}(100 {\rm ~GeV})$. If the LSP 
is higgsino-like, then $X_{\tau}\gtrsim20 {\rm ~TeV}$ (which is required for 
enhancing the $\hgamgam$ by at least 20\%) would require values of 
$\tan\beta\gtrsim200$ (unless $|A_{\tau}|\gtrsim20 {\rm ~TeV}$) leading 
to non-perturbative bottom and tau Yukawa couplings. Universal gaugino 
masses at the GUT scale are not compatible with a gaugino-like LSP with 
mass of about 100 GeV. The reason is that universal gaugino masses lead 
to the relation $M_1:M_2:M_3\approx1:2:6$ at the EW scale, as a 
consequence of the one-loop 
RGEs\footnote{
 The pattern of gaugino masses at the EW scale can be altered 
 in the presence of extreme hierarchy between the $A$-terms and gaugino 
 masses, $|A_0/M_{1/2}|>\mathcal{O}(100)$ because in such a case two-loop 
 effects in the RGEs for gaugino masses are non-negligible. See e.g.\ 
 Ref.~\cite{BaerDM} for an example of such a scenario.
},
which would require gluino mass to be smaller than about 
$\mathcal{O}(600 {\rm ~GeV})$ ($\mathcal{O}(300 {\rm ~GeV})$) for a  
bino (wino) LSP. Such light gluinos are excluded by the 
LHC.\footnote{
 Gluino could be so light only if the SUSY spectrum is extremely 
 compressed \cite{compressed_Dreiner} which is not the case with the LSP 
mass around 100 GeV.} 
The lower limit on the gluino mass varies depending on the features of the 
SUSY spectrum. Nevertheless, gluino mass below about 1.2 TeV is generically 
excluded even if the first-generation squarks are much heavier than the 
gluino \cite{ATLAS_gluino_bb,CMS_gluino_bb}, and the bound gets stronger 
with smaller masses of the first-generation squarks. The gluino mass above 
1.2 TeV together with a gaugino-like LSP with mass below about 100 GeV 
imply the following condition for the gaugino masses at the GUT scale:
\begin{equation}
\label{gaugino_cond}
 \left| c_1 \right| \equiv \left|\frac{M_1}{M_3}\right|\lesssim\frac{1}{2} 
\qquad {\rm or} \qquad  
 \left| c_2 \right| \equiv \left|\frac{M_2}{M_3}\right|\lesssim\frac{1}{4} \,,
\end{equation}
where we introduced parameters $c_1$ and $c_2$ defined as the GUT scale 
ratios of the bino and wino masses to the gluino mass. 
We should emphasize that the condition (\ref{gaugino_cond}) was obtained without assuming Yukawa unification so it is valid in any (R-parity conserving) MSSM model and should be valid also in many MSSM extensions such as NMSSM.
A more detailed discussion of ranges of $c_i$ values leading to enhanced 
$\hgamgam$ rates and compatible with the experimental constraints 
and unification of the Yukawa couplings is 
presented in the next Section\footnote{
It was shown very recently \cite{AkulaNath} that an enhanced $\hgamgam$ 
rate can be accommodated for a specific case of $c_1=c_2=1/10$,
however without imposing the Yukawa coupling unification. 
More general scenarios have not been studied yet.}.

\section{Top-bottom-tau Yukawa unification and enhanced Higgs diphoton rate}
\label{sec:YU}

We now would like to go one step further and ask the question whether it is
possible to obtain enhanced $\hgamgam$ rates in SO(10) models predicting 
top-bottom-tau Yukawa unification. Large values of 
$\tan\beta\sim\mathcal{O}(50)$ necessary for substantial stau 
mixing are predicted in such models. 
On top of the condition (\ref{gaugino_cond}) for enhanced 
$\hgamgam$ rate, there are also conditions for the MSSM parameters 
coming from the assumption of SO(10) GUT symmetry group and 
top-bottom-tau Yukawa unification which we discuss in the following.

\subsection{Top-bottom-tau Yukawa unification and REWSB}
\label{subsec:YU}

It is well known that in models with top-bottom-tau Yukawa unification 
proper REWSB is endangered because of the RGE effects of large bottom and 
tau Yukawa couplings which tend to make $M_A^2$ tachyonic \cite{YUreview}. 
In particular, REWSB is incompatible with top-bottom-tau Yukawa unification 
in CMSSM \cite{Carena}. In order to solve this problem some non-universalities 
in the soft scalar masses at the GUT scale need to be introduced 
\cite{Olechowski}. The pattern of soft scalar masses consistent 
with SO(10) gauge symmetry is given by:
\begin{align}
\label{scalarDterm}
&m_{H_d}^2=m_{10}^2+2D\,, \nn[4pt]
&m_{H_u}^2=m_{10}^2-2D\,, \nn[4pt]
&m_{Q,U,E}^2=m_{16}^2+D\,, \nn[4pt]
&m_{D,L}^2=m_{16}^2-3D\,, 
\end{align} 
where  $D$ parametrizes the size of a U(1) $D$-term contribution to soft 
scalar masses which generically arises in an effective theory below the GUT 
scale when SO(10) gauge symmetry (which has rank bigger than that of 
the SM gauge group) is spontaneously broken to its SM subgroup 
(or other subgroup with 
the rank smaller than that of SO(10)) \cite{Dterm}. The coefficients in 
front of $D$ are fixed by charges under the broken U(1). It has been shown 
that for $D>0$, $m_{10}>m_{16}$ and universal other soft terms proper 
REWSB is consistent with top-bottom-tau Yukawa unification \cite{Murayama}. 

The crucial role in top-bottom-tau Yukawa unification is played by the 
sign of $\mu$ because it controls the sign of the dominant finite 
SUSY threshold corrections to the bottom mass \cite{Hall,Carena,Pierce}: 
\begin{equation}
\label{mbsusycorr}
\left(\frac{\delta m_b}{m_b}\right)^{\rm finite}
\approx\frac{g_3^2}{6\pi^2}\mu m_{\tilde g}\tan\beta \, 
I(m_{\tilde b_1}^2,m_{\tilde b_2}^2,m_{\tilde g}^2)
+\frac{h_t^2}{16\pi^2}\mu A_t\tan\beta \, 
I(m_{\tilde t_1}^2,m_{\tilde t_2}^2,\mu^2) \,,
\end{equation}
where the loop integral $I(x,y,z)$ is defined e.g.\ in the appendix 
of \cite{Hall}. 
Bottom-tau Yukawa unification requires the finite correction to be 
negative with the magnitude between 10 and 20\% \cite{Wells_yuk}. 
The first term in eq.~(\ref{mbsusycorr}) comes from the gluino-sbottom 
contribution and is the dominant one over the most part of the parameter space 
so Yukawa unification strongly prefers $\mu<0$. Top-bottom-tau Yukawa 
unification is also possible for $\mu>0$ \cite{Blazek} but this requires very heavy 
scalars with $m_{16}$ exceeding $\mathcal{O}(20 ~{\rm TeV})$, and even 
larger $A$-terms, \cite{Baer_heaviergluino,YU_Rabynew} so corresponding 
fine-tuning of electroweak symmetry breaking is very big, much bigger than 
the fine-tuning imposed on the MSSM by the measured Higgs mass of 125 GeV. 
Moreover, staus in models with positive $\mu$ also tend to be heavy. 
The reason is that in those models gluinos have to be light and sbottoms 
very heavy in order to suppress the gluino-sbottom correction to the 
bottom mass (which has the wrong sign for positive $\mu$). The masses of 
staus cannot be much smaller than the masses of sbottoms because they have 
common value at the GUT scale and their RG evolution (which is dominated 
by the RGE terms proportional to the Yukawa couplings) is similar. 
It was found in Ref.~\cite{YU_Rabynew} that for positive $\mu$ staus are 
even heavier than sbottoms and that the properties of the 125 GeV Higgs 
are almost the same as those of the SM Higgs.
In contrast, for negative $\mu$ top-bottom-tau Yukawa unification is possible also for heavy gluino. As a result, staus can be much lighter than stops and sbottoms because RG contribution from gauginos to squarks may be large while the one to sleptons small.
For the above reasons in 
this paper we consider only $\mu<0$.

\subsection{Non-universal gaugino masses}
\label{subsec:Rgam}

As explained in Section \ref{sec:Rgam}, non-universal gaugino masses are necessary for enhancing $\hgamgam$ rate because
the experimental data from the LHC set lower bounds on the gluino mass. 
There is also another reason why gluinos should be rather heavy 
(what implies necessity of non-universal gaugino masses at the GUT scale). 
It is directly related to the requirement of Yukawa unification 
and enhanced $\Gamma(\hgamgam)$. The condition of $b$-$\tau$ Yukawa 
unification introduces rather strong correlation between $\mu$ and the 
gluino mass. Since $b$-$\tau$ Yukawa unification requires the finite 
threshold correction to the bottom mass to be below about 20\%, one 
can set the following bound on the ratio of the gluino mass and $\mu$  
\cite{YUreview}:
\begin{equation}
 \label{mubound}
\frac{m_{\tilde g}}{|\mu|}\gtrsim 2.5 \,,
\end{equation} 
where we used eq.~(\ref{mbsusycorr}) with the chargino-stop contribution 
neglected, $\tan\beta\approx50$ and assumed that gluino is heavier than 
the heaviest sbottom which is usually the case unless $M_3\ll m_{16}$. 
Big enhancement of $\hgamgam$ rate requires big $|\mu|$, e.g.\ 
for $m_{\tilde{\tau}_1}=100{\rm ~GeV}$ and $\tan\beta\approx50$ the 
enhancement by 20\% requires $|\mu|\gtrsim500\,$~GeV so the bound 
(\ref{mubound}) implies that the gluino mass has to be at least about 
1.2 TeV. Accidentally, this lower bound on the gluino mass coincides with 
the experimental bound so it leads to the same constraint 
(\ref{gaugino_cond}) on the gaugino masses at the GUT scale.

Even if gaugino masses at the GUT scale are such that the LSP is 
neutral when the lightest stau mass is around 100 GeV it is not 
guaranteed that the $\gamma\gamma$ rate is substantially enhanced. 
As explained in Section \ref{sec:Rgam}, also the second stau should 
be relatively light because the stau mixing should be close to 
the maximal (i.e.\ $\theta_{\bar{\tau}}\approx\pi/4$). 
For a fixed value of the off-diagonal term in the stau mass matrix, 
$X_\tau$, the stau mixing is maximized for $m_L\approx m_E$ at the EW scale. 
Due to the $D$-term splitting of the scalar masses (\ref{scalarDterm}) 
$m_L$ is smaller than $m_E$ at tree level. On the other hand, 
RG running of $m_L$ and $m_E$ is significantly different. 
In particular, the negative RGE contribution proportional to 
a large $\tau$ Yukawa coupling is two times larger for $m_E^2$ than 
the corresponding contribution to $m_L^2$ so this effect can diminish 
the initial splitting between $m_L$ and $m_E$. On top of that, the 
electroweak gaugino masses also contribute differently to $m_L$ and 
$m_E$ and this effect strongly depends on the values of $c_1$ and $c_2$. 
Therefore, the analysis of the $\gamma\gamma$ rate requires  
a careful numerical treatment.

In the following we focus on a numerical scan of the parameter space 
of a model with $D$-term splitting of scalar masses (\ref{scalarDterm}), 
universal trilinear coupling $A_0$ (justification for these 
assumptions will be discussed in the next Section), 
and arbitrary gaugino masses parametrized by $M_3$, $c_1$ and $c_2$. 
There are eight free parameters altogether so a homogeneous scan of 
the parameter space would not be very efficient. Therefore, we use 
Markov Chain Monte Carlo techniques to sample the parameter space. 
More precisely, we adopt a similar procedure to that proposed in 
Ref.~\cite{BaerDM} which makes use of the Metropolis-Hastings algorithm 
\cite{metropolis1,metropolis2}. We consider the following ranges of 
parameters:
\begin{align}
 &m_{16}\in\left(0, 10\ {\rm TeV}\right) \,, \nn
 &M_3 \in\left(0, 10\ {\rm TeV}\right) \,, \nn
  &\frac{m_{10}}{m_{16}}\in\left(0, 10\right) \,, \nn
&\frac{A_0}{m_{16}}\in \left(-3, 3\right) \,, \nn
&\frac{D}{m_{16}^2}\in \left(0, \frac{1}{3}\right) \,.
\end{align}
We also reject values of $D$ that lead to a negative value of any of 
the soft scalar squared masses (\ref{scalarDterm}) at the GUT scale. 
On the other hand, for $c_1$, $c_2$ and $\tan\beta$ we do not specify 
ranges over which they are scanned.

In order to quantify the goodness of top-bottom-tau Yukawa unification 
we introduce the following quantity:
\begin{equation}
 R\equiv\left.\frac{\max\left(h_t,h_b,h_{\tau}\right)}
{\min\left(h_t,h_b,h_{\tau}\right)}\right|_{\rm GUT} \,.
\end{equation} 
We use SOFTSUSY \cite{softsusy} to solve the 2-loop renormalization 
group equations and calculate the MSSM spectrum. For every randomly 
generated point in the parameter space we demand proper REWSB and 
one of the neutralinos being the LSP. We also reject points that do 
not satisfy the vacuum metastability condition (\ref{stability_cond}). 
We calculate the thermal relic abundance of the lightest neutralino, 
as well as BR$(b\to s\gamma)$, BR$(B_s\to\mu^+\mu^-)$ and $(g-2)_{\mu}$ 
using MicrOmegas \cite{Micromega}. We apply the following constraints:
\cite{bsg,bsgth,bsgth2,Bsmumu_comb,Bsmumu_th,WMAP,Planck}
\begin{eqnarray}
\label{exp_constr}
& 2.52\cdot10^{-4}<{\rm BR}(b\to s\gamma)<4.34\cdot10^{-4} \,, \nn 
& 1.5\cdot10^{-9}<{\rm BR}(B_s\to\mu^+\mu^-)<4.3\cdot10^{-9} \,, \nn 
& \Omega_{\rm DM}h^2<0.13  \,.
\end{eqnarray}
For ${\rm BR}(b\to s\gamma)$ we use the $2\sigma$ experimental 
constraint combined in quadrature with the theoretical uncertainty 
of $4\cdot10^{-5}$. Note that the computation of ${\rm BR}(b\to s\gamma)$ 
in the MSSM is completed at the NLO \cite{bsgth2}, while in the SM at 
the NNLO \cite{bsgth}. Moreover, the NNLO corrections shift the NLO 
result in the SM \cite{bsgSM_NLO} by about $4\cdot10^{-5}$ so we use 
the value of this shift as an estimate of the theoretical error in MSSM. 
The theoretical uncertainty for ${\rm BR}(B_s\to\mu^+\mu^-)$ 
\cite{Bsmumu_th} is still much smaller than the experimental one so 
we use for this observable the $2\sigma$ limit obtained by combining \cite{Bsmumu_comb} the CMS \cite{Bsmumu_CMS} and LCHb \cite{Bsmumu_LHCb} results.
We demand that only the upper bound 
on $\Omega_{\rm DM}h^2$ is satisfied but, as we shall discuss later,  
also the lower bound can be satisfied in some circumstances. 
The relevant lower mass limits on the MSSM 
particles\cite{LEPstau,CMS_MA,Atlas_discovery,CMS_discovery}:
\begin{align}
\label{exp_masslimits}
90\, {\rm GeV} < \,\,& m_{\tilde{\tau}}\,\,, \nn
103.5\, {\rm GeV} <\,\,  & m_{\tilde{\chi}^{\pm}}\, \,, \nn
750\, {\rm GeV}<\,\,& m_A \,\,, \nn
123\, {\rm GeV}<\,\, & m_h<128\, {\rm GeV}
\end{align}
are also applied. 
Experimental lower mass limit on $m_A$ depends on $\tan\beta$ but we fix it 
to a constant value since the assumption of top-bottom-tau Yukawa unification 
constrains $\tan\beta$ to be between 40 and 50, where the limit of 750 GeV 
is a good approximation. 
The theoretical uncertainty in the prediction of the Higgs mass, calculated 
by SOFTSUSY at two-loop level, is about 3 GeV \cite{Allanach_higgs} so we 
assume that the Higgs mass between 123 and 128 GeV is consistent with the 
experimentally measured value of about 125.5 GeV. In practice, we found 
that only the lower bound on the Higgs mass constrains the parameter space.  
It was argued in Ref.~\cite{Higgs3loop} that dominant three-loop 
corrections to the Higgs mass are positive with the magnitude 
up to 3 GeV. These results 
strengthen our assumption that points for which SOFTSUSY gives the Higgs 
mass of 123 GeV are compatible with the experimental data. 
We found that other lower limits on sparticle masses from direct LHC 
searches do not impose any additional constraints on the model. 

We also use the quantity $R_{\gamma\gamma}$ defined as the predicted 
signal strength in the $\gamma\gamma$ channel normalized to the 
corresponding SM prediction:
\begin{equation}
  R_{\gamma\gamma}
\equiv
\frac{\sigma(gg\to h)\times{\rm BR}(h\rightarrow\gamma\gamma)}
{\sigma(gg\to h)^{\rm SM}\times{\rm BR}(h\rightarrow\gamma\gamma)^{\rm SM}}\,.
\end{equation}
The strong LHC constraints on $m_A$ push the model to the decoupling 
region of the MSSM so the fermion and gauge boson couplings are almost 
the same as in the SM. In consequence, both the production cross-section 
and the total decay width of the Higgs are practically the same as in the SM. 
In principle, light stops or sbottoms could modify the Higgs production 
cross-section and decay width into $\gamma\gamma$ but we found in our 
numerical analysis that stops and sbottoms are relatively heavy and 
such effects are negligible. Therefore, any deviations of 
$R_{\gamma\gamma}$ from one in this model are due to 
light and strongly-mixed staus. We numerically compute $R_{\gamma\gamma}$ using the formulae collected in Subsection \ref{subsec:staus}.

\begin{figure}[t!]
  \begin{center}
    \includegraphics[width=0.49\textwidth]{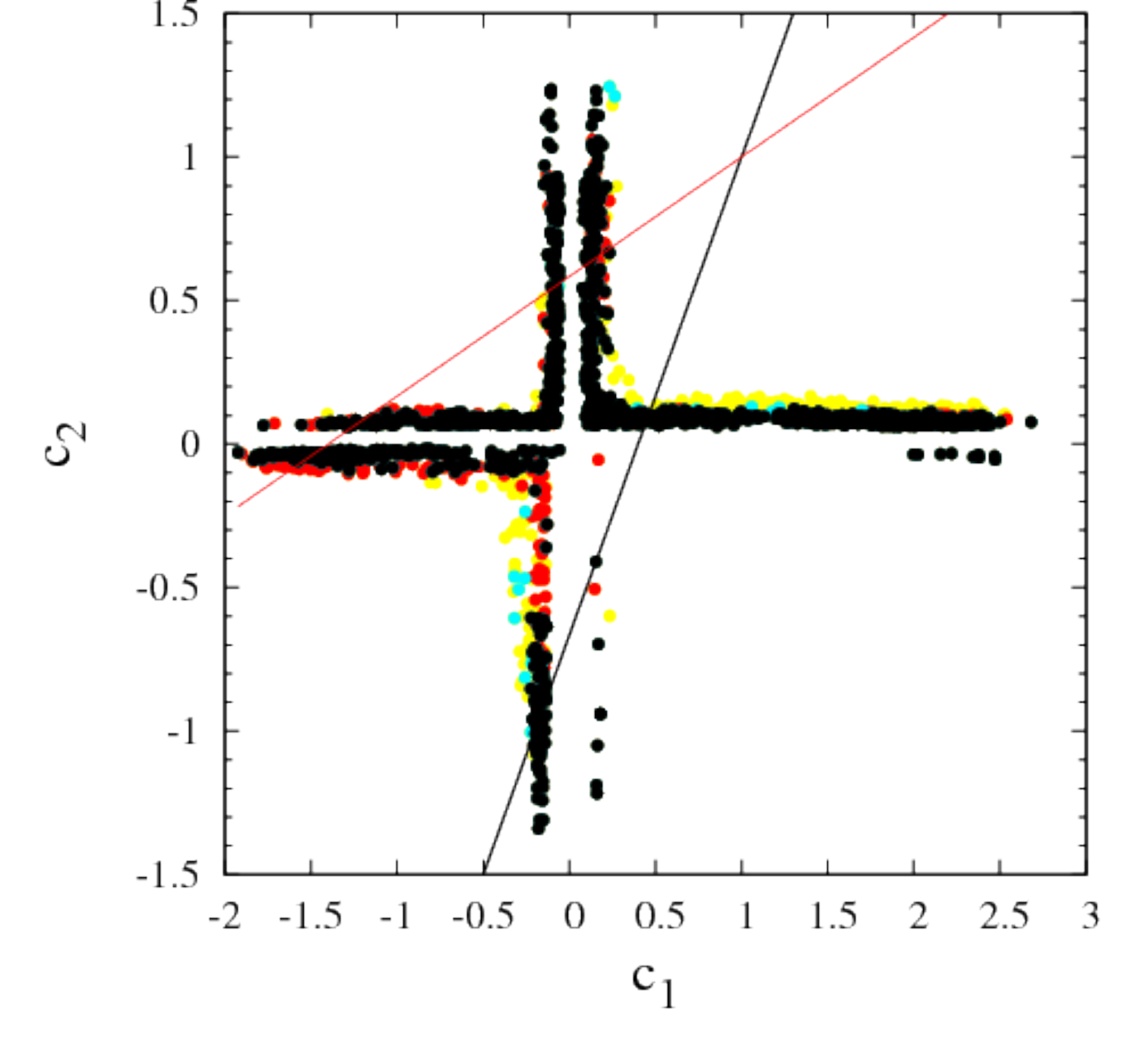}
    \caption{Points with $R_{\gamma\gamma}>1.1$ and $R<1.1$ in the 
$c_1$--$c_2$ plane. Black points satisfy all the constraints in 
(\ref{exp_constr}) and (\ref{exp_masslimits}). Blue (red) points 
violate $b\to s \gamma$ ($B_s\to \mu^+\mu^-$). Yellow points 
violate both $b\to s \gamma$ and $B_s\to \mu^+\mu^-$. Black 
(red) line corresponds to gaugino masses generated by a mixture 
of the singlet and ${\bf{24}}$ $F$-term (in mirage mediation).
    }
    \label{fig:c1c2}
  \end{center}
\end{figure}

In Figure \ref{fig:c1c2} we present the results of our scan in the 
$c_1$--$c_2$ plane. Only points that correspond to $R<1.1$ (Yukawa 
unification better than 10\%) and $R_{\gamma\gamma}>1.1$ (enhancement 
of the $\gamma\gamma$ rate by at least 10\%) are shown. Black points 
satisfy all the constraints mentioned before, including the 
metastability bound (\ref{stability_cond}). For blue (red) points 
$b\to s \gamma$ ($B_s\to \mu^+\mu^-$) constraint is relaxed, while yellow 
points violate both $b\to s \gamma$ and $B_s\to \mu^+\mu^-$. Note that 
all points have either $c_1$ or $c_2$ small, in agreement with our 
qualitative constraint (\ref{gaugino_cond}). However, there are very few points in the quadrant with $c_1>0$ and $c_2<0$. In this quadrant of the $c_1$--$c_2$ plane the Yukawa-unified solutions with enhanced $\hgamgam$ rate are characterized by a light pseudoscalar Higgs with mass typically below the experimental lower bound \cite{CMS_MA}. The lightness of the CP-odd Higgs in this region of parameter space follows from the assumption of Yukawa unification. Without imposing this assumption enhanced $\hgamgam$ rate can be easily obtained also in the quadrant with $c_1>0$ and $c_2<0$.

The scan with arbitrary gaugino masses is very useful for illustrative 
purposes but it is more interesting to focus on specific models of SUSY 
breaking that predict some patterns of gaugino masses. One example of 
such pattern is mirage mediation \cite{mirage1}-\cite{mirage7} which 
predicts  
\begin{equation}
M_a = M(\rho + b_a g_a^2)\,,
\label{mirage}
\end{equation}
where $M$ and $\rho$ are free parameters, 
$b_a = (33/5, 1, -3) \; {\rm for} \; a = 1, 2, 3$ 
and $g_a$ are the gauge coupling constants 
(which in our model are assumed to unify at the GUT scale). 
In this case the relation between $c_1$ and $c_2$ is fixed:
\begin{equation}
 c_2=\frac{5c_1 + 7}{12} 
\end{equation}
and corresponds to the red line in Figure \ref{fig:c1c2}. The line 
intersects regions in which top-bottom-tau Yukawa unification and 
enhanced $\hgamgam$ rate can be obtained. 
This indicates that Yukawa unification with enhanced $\gamma\gamma$ 
rate can be realized in some mirage mediation scenario. It is an 
indication and not a proof because the points shown in 
Figure \ref{fig:c1c2} do not correspond exactly to any mirage 
mediation model. Those points were obtained assuming universal 
trilinear soft terms while mirage mediation models predict 
trilinear terms with non-universalities correlated with the 
non-universalities of the gaugino masses (\ref{mirage}).
Thus, to draw any firm conclusions about such models it is necessary 
to perform separate calculations dedicated to each of them. 
Our preliminary results 
show that top-bottom-tau Yukawa unification and 
enhanced $\hgamgam$ rate indeed can be obtained in some 
mirage mediation 
models.\footnote{
Yukawa unification in ``effective'' mirage mediation was 
discussed in a recent Ref.\ \cite{YUmirage}, however, 
neglecting the necessary non-universalities of the $A$-terms 
and without any discussion of $\Gamma(\hgamgam)$.
}
 The full results will be presented 
elsewhere. In the present paper we concentrate on another 
class of models.

Non-universal gaugino masses are also possible in GUT models with 
pure gravity mediation provided that the SUSY breaking $F$-term 
belongs to an appropriate non-singlet representation of the 
unifying gauge group. In order to see this, notice first that the 
gaugino masses in supergravity can arise from the following dimension 
five operator:
\begin{equation}
\label{gaugino_op}
{\cal L}\supset -\frac{F^{ab}}{2 M_{\rm Planck}}
\lambda^a \lambda^b + {\rm c.c.}\,,
\end{equation}
where $\lambda^a$ are the gaugino fields and the resulting gaugino mass 
matrix is $\frac{\langle F^{ab}\rangle}{M_{\rm Planck}}$. 
Non-zero gaugino masses require that the vacuum 
expectation value of the relevant $F$-term, $\langle F^{ab}\rangle$, 
transforms as the singlet of the SM gauge group and, in order 
to make the term (\ref{gaugino_op}) invariant under the GUT group, 
as any of the representations present in the symmetric part of 
the direct product of the two adjoint representations, which 
for SO(10) are:
\begin{equation}
({\bf 45} \times {\bf 45} )_{\rm S} 
=  {\bf 1} + {\bf 54} + {\bf 210} + {\bf 770} \,.
\label{45x45}
\end{equation} 
If SUSY is broken by an $F$-term transforming as a non-singlet 
representation of SO(10), gaugino masses are not universal. 
Complete classification of non-universal gaugino
masses for SO(10) and its subgroups can be found in Ref.~\cite{Martin}.

Particularly interesting is the case when the SUSY breaking $F$-term 
transforms as ${\bf 24}$ of SU(5) $\subset$ SO(10) for which the 
gaugino masses have the following pattern:
\begin{equation}
\label{gaugino24}
 M_1:M_2:M_3 = -\frac{1}{2}:-\frac{3}{2}:1\,.
\end{equation}
The negative value of $M_2$ (with respect to $M_3$) is preferred 
from the phenomenological point of view because for $\mu<0$ such 
values make the SUSY contribution to $(g-2)_{\mu}$ positive. 
Moreover, for $\mu<0$ and $M_2<0$ 
the chargino-stop contribution to $b\to s \gamma$ is smaller than 
for $M_2>0$ so the tension with this observable is relaxed. 
Top-bottom-tau Yukawa unification in a model with gaugino masses 
generated by the SUSY breaking $F$-term in ${\bf 24}$ of SU(5) 
$\subset$ SO(10) and the soft scalar masses (\ref{scalarDterm}) 
with $D$-term splitting was investigated in Ref.~\cite{bop_F24} 
with a special emphasize on the constraints from $(g-2)_{\mu}$ 
and $b\to s \gamma$, while the LHC constraints on that model 
from the Higgs and SUSY searches were studied in Refs.~\cite{mbks,YUreview}.

\begin{figure}[t!]
  \begin{center}
    \includegraphics[width=0.49\textwidth]{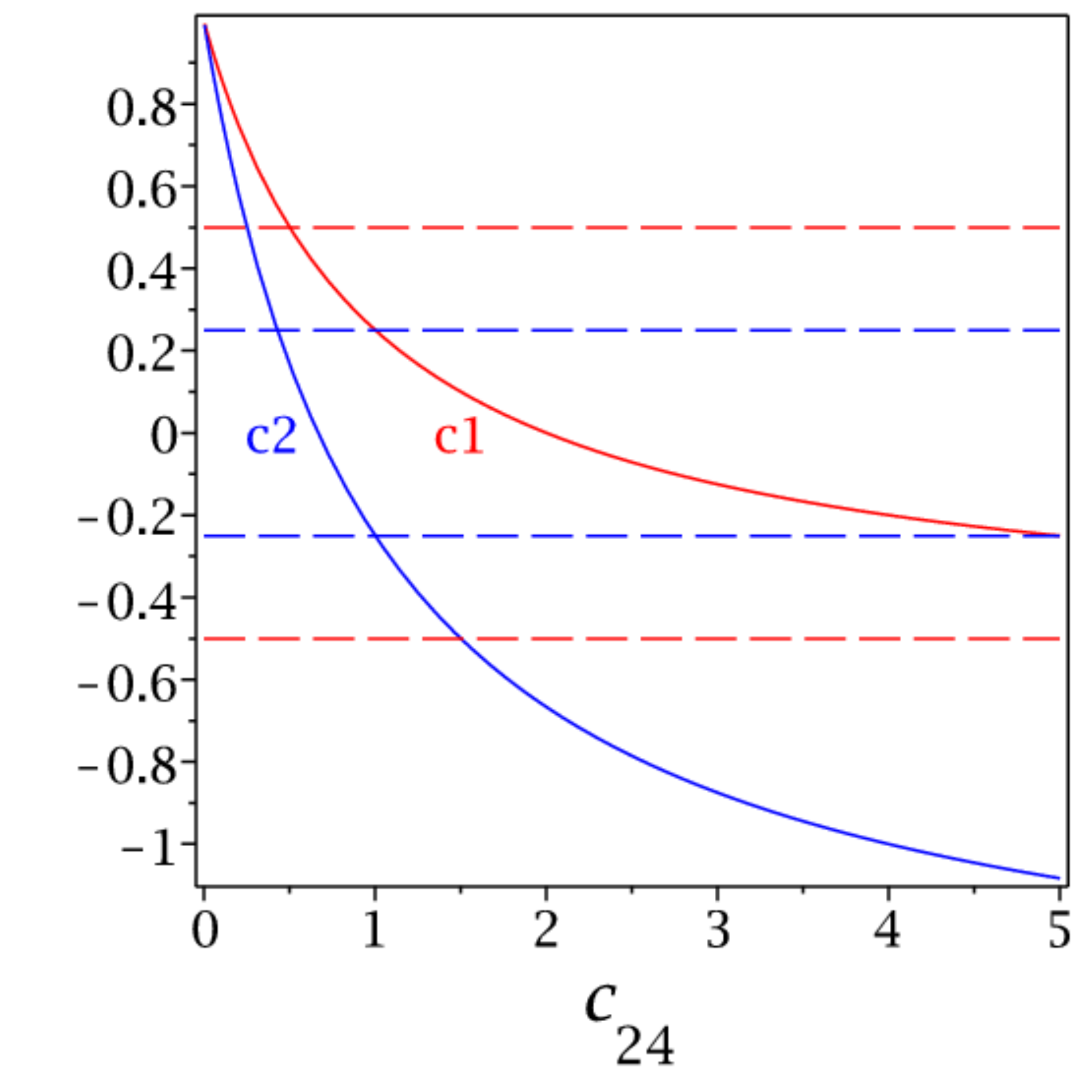}
    \caption{ $c_1$ (red) and $c_2$ (blue) as functions of $c_{24}$. 
The dashed lines correspond to $|c_1|=1/2$ and $|c_2|=1/4$.
    }
    \label{fig:ci_c24}
  \end{center}
\end{figure}

The gaugino mass pattern (\ref{gaugino24}) is interesting but still not 
suitable to allow for substantial enhancement of $\hgamgam$ rate. 
It gives $c_1=-0.5$ and $c_2=-1.5$ which are too big values, as can be 
seen in Figure \ref{fig:c1c2}. This problem may be solved when  
supersymmetry is broken by more than just one $F$-term. The simplest 
possibility is to consider two such $F$-terms, one transforming as 
${\bf 24}$ of SU(5) $\subset$ SO(10) and second transforming 
as the gauge singlet. If both such $F$-terms have non-zero VEVs  
the gaugino masses can be parametrized in the following way:
\begin{align}
\label{gauginoF1F24}
M_1&=M_3^{(1)}-\frac{1}{2}M_3^{(24)}
\equiv \frac{1-\frac{1}{2}c_{24}}{1+c_{24}}M_3  \,, \nn[1pt]
M_2&=M_3^{(1)}-\frac{3}{2}M_3^{(24)}
\equiv \frac{1-\frac{3}{2}c_{24}}{1+c_{24}}M_3  \,, 
\end{align} 
where $c_{24}\equiv M_3^{(24)} / M_3^{(1)}$ is the ratio of the ${\bf 24}$ 
and singlet $F$-term contributions to the gluino mass. 
In this case the relation between $c_1$ and $c_2$ is given by:
\begin{equation}
 c_2=\frac{5c_1 - 2}{3} 
\end{equation}
and corresponds to the black line in Figure \ref{fig:c1c2}. This line 
intersects several regions with good Yukawa unification and enhanced 
$\hgamgam$ rate. A more detailed analysis 
of these regions will be presented in the next Section.

In order to get 
a feeling how the ratios of the gaugino masses,  $M_1/M_3$ and $M_2/M_3$,
depend on $c_{24}$ we plot them in Figure \ref{fig:ci_c24}.
Notice that $c_{24}$ has to be positive and larger 
than about 0.4 in order to get $|c_1|\lesssim1/2$ or $|c_2|\lesssim1/4$. 
Moreover, for $c_{24}$ below (above) about 0.9 the LSP is dominated by 
the wino (bino). Notice also that for $c_{24}<2/3$, $M_2>0$ so the  
contribution to $(g-2)_{\mu}$ from the chargino-sneutrino loop, 
which is typically a dominant SUSY contribution \cite{Moroi,Stockinger}, 
is negative and the discrepancy between the theoretical prediction 
and experimental result becomes even larger than in the SM.

\section{Model with SUSY broken by {\boldmath $F$}-terms in 1 
and 24 representations of SU(5) }
\label{sec:F1F24}

Phenomenological implications resulting from non-universal gaugino 
masses generated by a mixture of the singlet and ${\bf 24}$ $F$-term 
were investigated before \cite{F1F24King,F1F24Martin}. However, 
neither the impact of that assumption on Yukawa unification nor the 
$h\to \gamma\gamma$ rate has been considered so far. 
In this Section we study in detail implications of these assumptions. 
First of all, we should check how the assumed structure of the 
$F$-terms influences all the soft SUSY breaking terms. In the 
previous Section we concentrated on the gaugino masses assuming 
for simplicity that the trilinear terms are universal and the 
structure of the soft scalar masses is as given in eq.\ (\ref{scalarDterm}). 
We show in the following Subsection whether and when such 
simplifying assumptions can be justified.

\subsection{(Non)universalities of other soft terms}
\label{subsec:SoftTerms}

${\bf 24}$ of SU(5) appears in each of the three non-singlet 
representations of SO(10) in the symmetric part of the 
product ${\bf 45} \times {\bf 45}$ given in eq.\ (\ref{45x45}).
The pattern of the gaugino masses is the same for each 
of these SO(10) representations. This is not true 
for the soft scalar masses and trilinear terms.

Soft scalar masses terms arising in supergravity from 
dimension six operators have the following structure
\begin{equation}
\label{scalar_op}
\frac{\left<F\, F\right>^{ij}}{M_{\rm Planck}^2}
\phi_i^\dagger \phi_j\,,
\end{equation}
where $\phi_{i,j}$ are scalar fields. In order to contribute 
to the scalar masses without breaking the SM gauge symmetry, 
the VEV of the product of $F$-terms in the 
above formula must transform as a singlet of the SM and 
some representation present in the product
\begin{equation}
{\bf {\overline{16}}} \times {\bf 16} 
=  {\bf 1} + {\bf 45} + {\bf 210} \,.
\label{16x16bar}
\end{equation} 
Any $F$-term transforming as a tensor representation ${\bf R}$ 
of SO(10) gives some universal contribution to the scalar 
masses because the singlet is always present in the product  
$({\bf R \times R})_{\rm S}$. In addition, some of them may give also 
non-universal contributions. This happens if the symmetric
part of the product ${\bf R \times R}$ contains ${\bf 45}$ 
or ${\bf 210}$. This is the case for the representations 
${\bf 210}$ and ${\bf 770}$ but not for 
${\bf 54}$.\footnote{
  The products ${\bf 54} \times {\bf 54}$ and 
  ${\bf 210} \times {\bf 210}$ are given in Ref.\ \cite{Slansky}.
  In the case of ${\bf 770} \times {\bf 770}$ one can 
  prove that it contains ${\bf 45}$ and ${\bf 210}$ 
  using the Young tableaux technique.}
Let us also note that for the representation ${\bf 210}$ additional contribution to the soft scalar masses may arise from the mixed product of the singlet and non-singlet $F$-terms because ${\bf 210}$ is present in both products, (\ref{45x45}) and (\ref{16x16bar}).

The soft trilinear terms, generated by dimension five 
operators, have the form
\begin{equation}
\label{A_op}
\frac{\left<F\right>^{ijk}}{M_{\rm Planck}}
\phi_i \phi_j\phi_k\,,
\end{equation}
and may arise for $F$-terms transforming as any of the representations 
appearing in the product
\begin{equation}
({\bf {16}} \times {\bf 16})_{\rm S} \times  {\bf 10}
=  {\bf 1} + {\bf 45} + {\bf 54} + {\bf 210} + {\bf 1050}\,.
\label{16x16x10}
\end{equation} 
Let us now discuss in turn each of the non-singlet representations 
present in the r.h.s.\ of eq.\ (\ref{45x45}). Representation 
{\bf 770} is absent in the r.h.s.\ of eq.\ (\ref{16x16x10})  
so the corresponding $F$-term does not generate any trilinear terms. 
$F$-terms transforming as {\bf 54} and {\bf 210} do generate soft 
trilinear terms. In fact, {\bf 210} leads to two kinds of such 
terms: universal and non-universal. This follows from the fact 
that there are two independent singlets in the product 
$({\bf {16}} \times {\bf 16})_{\rm S} \times  {\bf 10} \times {\bf 210}$.
There is only one contribution to the trilinear terms coming 
from $F$-terms transforming as {\bf 54}. A more detailed analysis 
shows that this contribution is universal.

The last results may be understood using the following argument.
In the only singlet in the product 
$({\bf {16}} \times {\bf 16})_{\rm S} \times  {\bf 10} \times {\bf 54}$
the part containing the Higgs multiplet and the $F$-term multiplet, 
${\bf 10} \times {\bf 54}$, transforms as {\bf 10}. Representation 
{\bf 10} may be understood as a vector of SO(10) while {\bf 54} 
as a symmetric matrix. Their product transforming as {\bf 10} is 
obtained by multiplying this vector with this matrix. $\left<F\right>$ 
must be a SM singlet so it is represented by a diagonal matrix. 
Moreover, this matrix is proportional to unit matrices in the 
subspaces corresponding to SU(3) and SU(2) subgroups of SO(10).
So, the only non-universality generated by {\bf 54} is that between 
trilinear terms involving Higgs doublets versus trilinear terms 
involving Higgs triplets. We neglect the latter assuming some 
mechanism for the Higgs doublet-triplet splitting.
Trilinear terms for the Higgs doublets alone are universal.

Let us summarize the above results. There are three possibilities 
in models with SUSY broken by two $F$-terms, one transforming as {\bf 1} 
and one as {\bf 24} of SU(5). Each case leads to the same pattern 
of the non-universalities in the gaugino masses, the one discussed 
in the previous Subsection. The patterns of other soft terms 
depends on the representation of SO(10) in which {\bf 24} of 
SU(5) is embedded. In the case of {\bf 54} of SO(10) all other
soft terms are universal at the GUT scale. Representation 
{\bf 210} leads to non-universality in soft scalar masses 
and trilinear terms. Finally, {\bf 770} gives non-universal 
scalar masses but universal trilinear terms.

\subsection{Numerical results}
\label{subsec:results}

The discussion in the previous Subsection shows that embedding 
${\bf 24}$ of SU(5) in ${\bf 54}$ of SO(10) is the most attractive 
possibility. We have shown that in such a case all soft trilinear 
terms have one common value, $A_0$, at the GUT scale. 
As for the soft scalar masses we assume the pattern given by 
(\ref{scalarDterm}). The $F$-term transforming as {\bf 54} does 
not give non-universalities in the scalar masses but the 
difference between $m_{10}$ and $m_{16}$ may be generated by the 
RGE running between the Planck scale and the GUT 
scale.\footnote{
  Using {\bf 210} or {\bf 770} instead of {\bf 54} would 
  lead to much more complicated pattern of scalar masses,
  splitting e.g.\ masses of sfermions belonging to one 
   {\bf 16} representation of SO(10).
}
Our choice is also the simplest because {\bf 54} is the smallest 
representation of SO(10) leading to non-universal gaugino masses.

Having defined the model we can investigate its properties. 
We perform similar numerical scan to the one described in the 
previous Section with the only difference that the gaugino masses 
are determined by (\ref{gauginoF1F24}) as a function of $c_{24}$ 
and $M_3$ (i.e.\ we scan along the black line in Fig.\ \ref{fig:c1c2}).

\begin{figure}[t!]
\begin{center}
\includegraphics[width=0.49\textwidth]{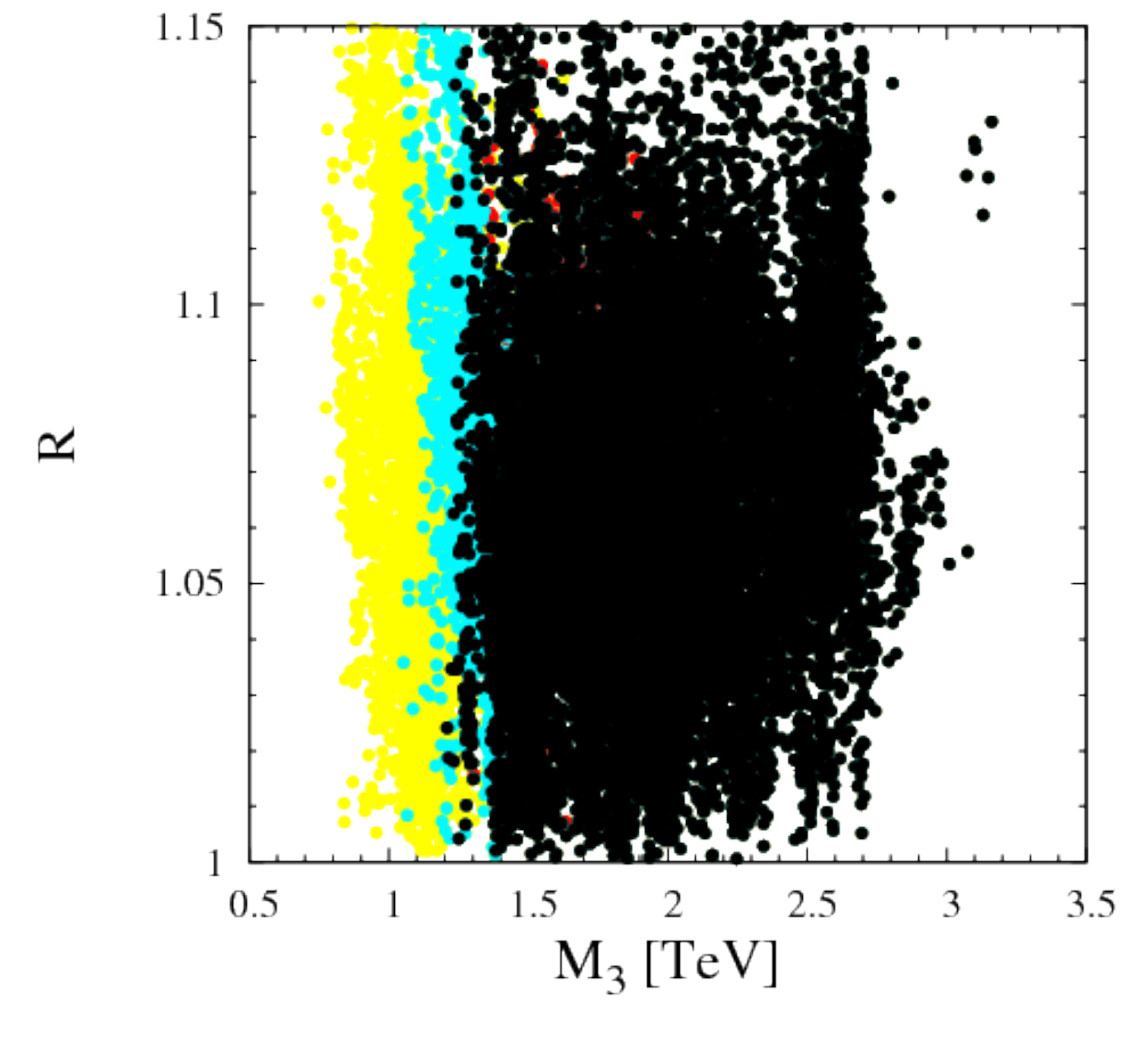}
\caption{Points with $R_{\gamma\gamma}>1.1$ in the $R$--$M_3$ plane. 
The colour coding is as in Figure \ref{fig:c1c2}.
\label{fig:R_M3}
} 
\end{center}
\end{figure}

\begin{figure}[t!]
\begin{center}
\includegraphics[width=0.49\textwidth]{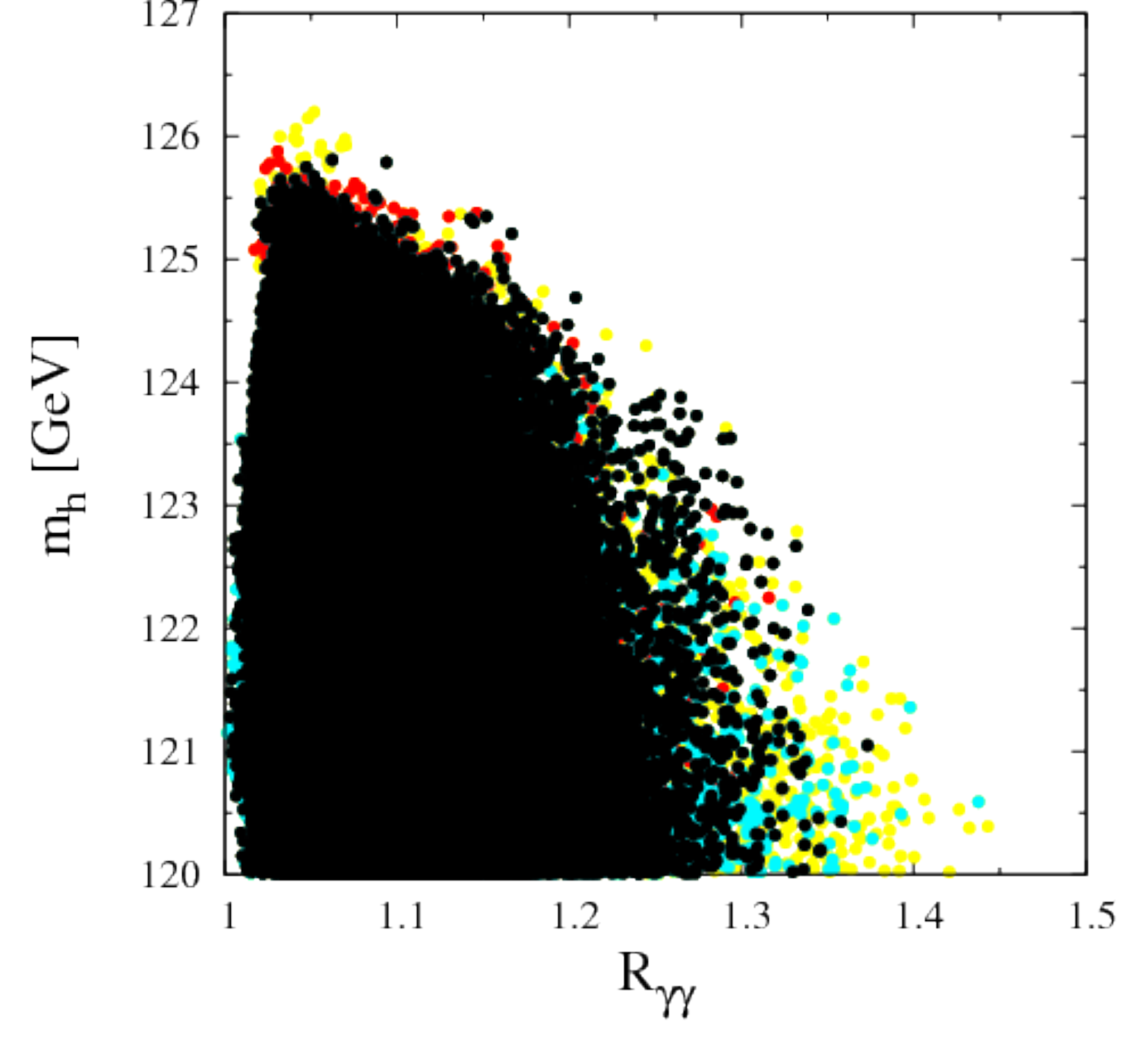}
\hfill
\includegraphics[width=0.49\textwidth]{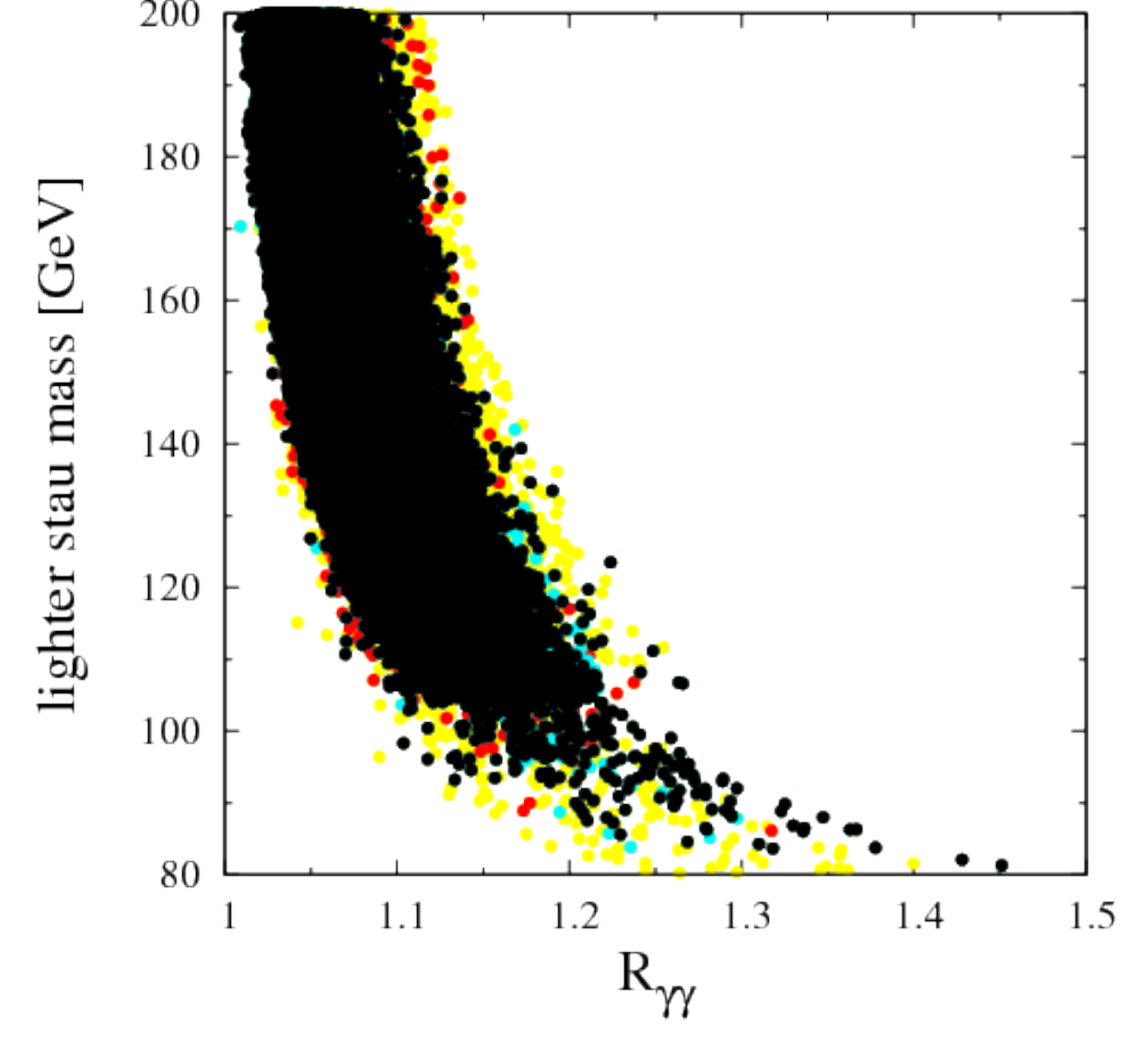}
\caption{Scatter plots of $m_h$ (left panel) and the 
lighter stau mass (right panel) versus 
$R_{\gamma\gamma}$ for the points with $R<1.1$. 
Points in the left panel were obtained without imposing 
the bound (\ref{exp_masslimits}) on the Higgs boson mass. 
The colour coding is the same as in Figure \ref{fig:c1c2}.
} 
\label{fig:mh_Rgam_mstau}
\end{center}

\end{figure} 

In Figure \ref{fig:R_M3} we present points from our scan characterized 
by $R_{\gamma\gamma}>1.1$ in the $R$--$M_3$ plane. It shows that  
$R_{\gamma\gamma}>1.1$ may be consistent with the experimental 
constraints provided that the gluino is heavy enough. 
The lower bound on $M_3$ could be 
somewhat relaxed if the constraints from $b\to s \gamma$ and 
$B_s\to\mu^+\mu^-$ were not taken into account. 
Moreover, the bound on the gluino mass 
depends very weakly on $R$ and even demanding the 
perfect top-bottom-tau Yukawa unification, i.e. $R\approx1$, 
is not an obstacle to get enhanced $h\to\gamma\gamma$ rate.

In the left panel of Figure \ref{fig:mh_Rgam_mstau} we plot the 
points with $R<1.1$ in the $m_h-R_{\gamma\gamma}$ plane. It can be seen 
from this plot that $h\to\gamma\gamma$ rate can be enhanced even by 30\%. 
Notice also that $R_{\gamma\gamma}$ is anti-correlated with the Higgs mass. 
This is because $R_{\gamma\gamma}$ grows when 
$X_{\tau}^2/(m_{\tilde{\tau}_1} m_{\tilde{\tau}_2})$ increases but this, 
in turn, implies that negative contribution to the Higgs mass from 
the stau sector grows as well \cite{Carena_stau}. The anti-correlation 
is weaker for smaller values of $m_h$ because too large values of 
$X_{\tau}^2/(m_{\tilde{\tau}_1} m_{\tilde{\tau}_2})$ are excluded by the 
constraints on the vacuum metastability.

The plot in the right panel of Figure \ref{fig:mh_Rgam_mstau} confirms 
our expectations showing that large enhancement of the $\gamma\gamma$ 
rate requires the lighter stau to have mass around 100 GeV. Taking the 
most conservative lower limit on the stau mass of 82 GeV even 40\% 
enhancement can be obtained. Note, however, that non-negligible 
enhancement do not require extremely light stau. For instance, 
$R_{\gamma\gamma}>1.1$ can be obtained for the lightest stau mass 
as large as about 200 GeV.

\begin{figure}[t!]
\begin{center}
\includegraphics[width=0.49\textwidth]{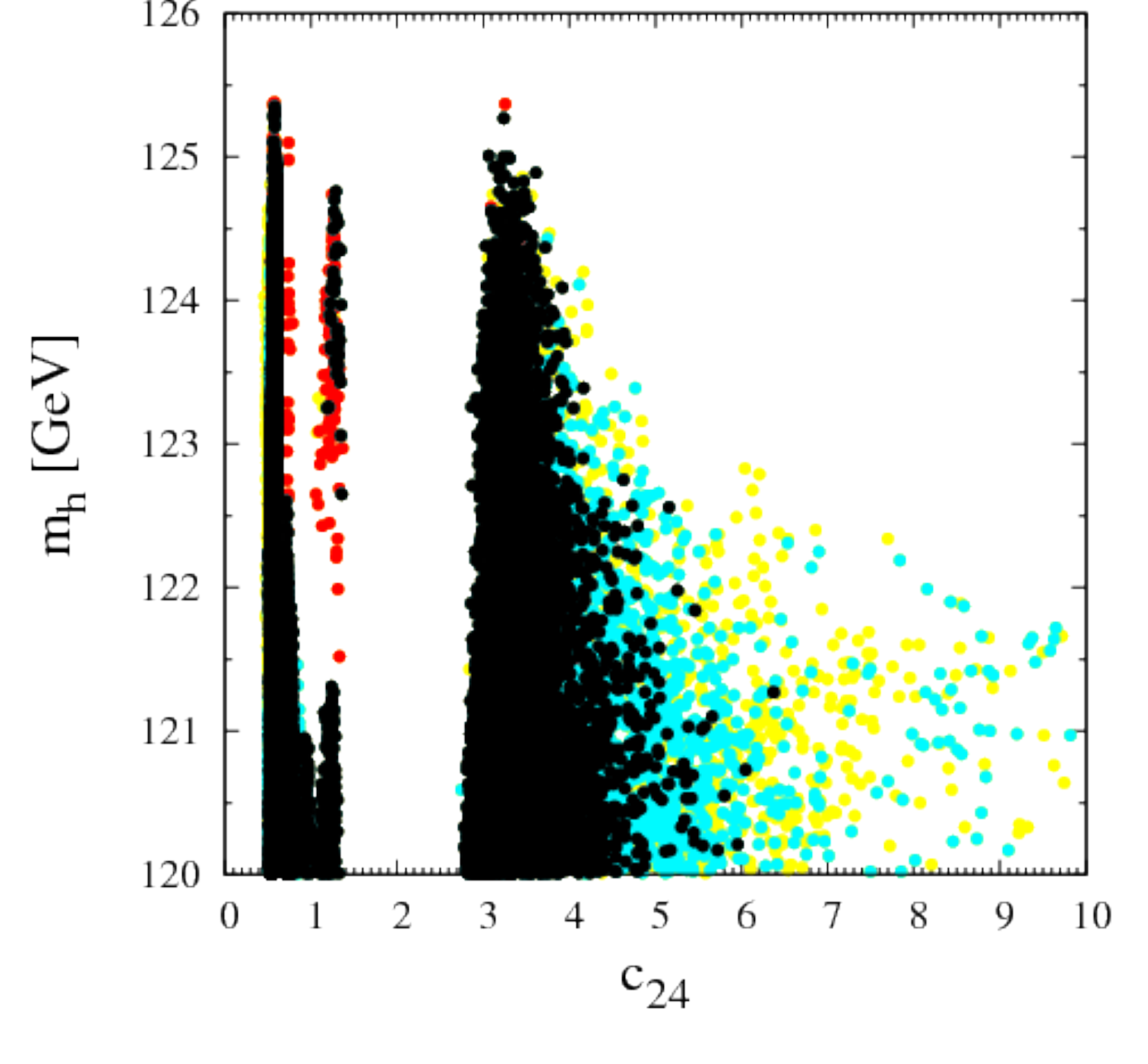}
\hfill
\includegraphics[width=0.49\textwidth]{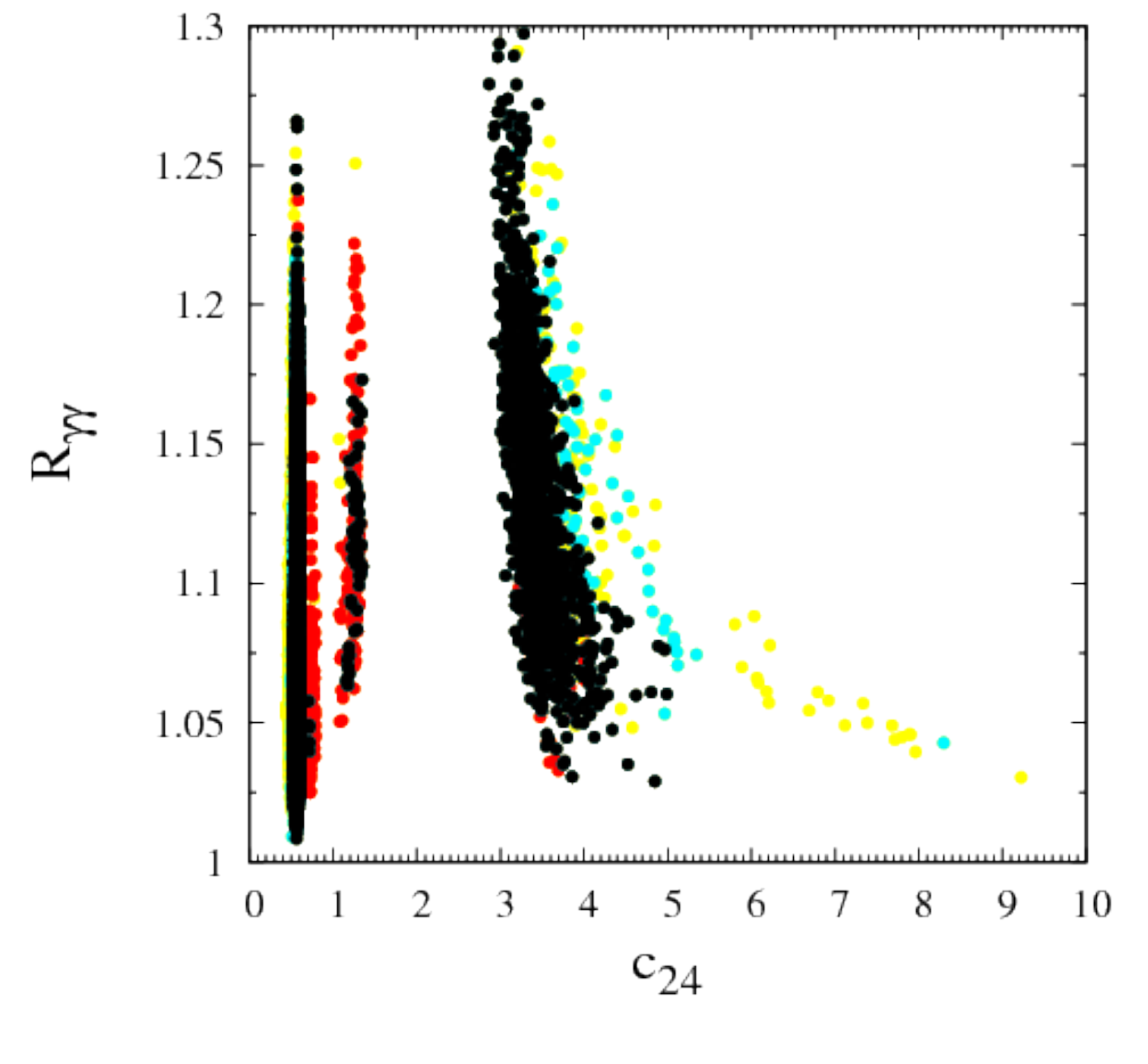}
\caption{
Scatter plots of $m_h$ (left panel) and $R_{\gamma\gamma}$ (right panel) 
versus $c_{24}$ for the points with $R<1.1$. 
Points in the left panel have $R_{\gamma\gamma}>1.1$ but they were obtained 
without imposing the bound (\ref{exp_masslimits}) on the Higgs boson mass. 
The colour coding is the same as in Figure \ref{fig:c1c2}.
} 
\label{fig:mhRgam_c24}
\end{center}
\end{figure} 

In Figure \ref{fig:mhRgam_c24} we present the dependence of the Higgs 
mass and $R_{\gamma\gamma}$ on $c_{24}$. In accord with our qualitative 
discussion in Subsection \ref{subsec:Rgam}, $m_h>123$ GeV and 
$R_{\gamma\gamma}>1.1$ is 
possible only if $c_{24}$ is positive and larger than about 0.5 but smaller 
than about 4 (5) when the $B$-physics constraints are (not) taken into 
account. However, not all values of $c_{24}$ in this range are possible 
due to various phenomenological constraints. Since for $c_{24}=2/3$ 
the wino mass vanishes, values of $c_{24}$ very close to this value 
are excluded by the LEP constraint on the chargino mass. For $c_{24}=2$ 
the bino mass vanishes so values of $c_{24}$ close to this value imply 
very light bino-like LSP with too large thermal relic abundance which 
cannot be reduced by stau coannihilation because the mass splitting 
between LSP and stau is too large. On the other hand, moving 
away from $c_{24}=2$ increases the bino mass so the mass splitting 
between LSP and stau gets smaller. For $c_{24}\lesssim1.4$ and 
$c_{24}\gtrsim3$ the mass splitting can be small enough to have 
the bino-like LSP relic abundance in agreement with observations. 
Finally, also some region around $c_{24}=1$ is excluded because 
the pseudoscalar Higgs mass turns out to be below 750 GeV there.
We should stress at this point that very light pseudoscalar Higgs in the region around $c_{24}=1$ is a consequence of the assumption of Yukawa unification. Without this assumption enhanced $\hgamgam$ rate can be obtained also in this region with the pseudoscalar Higgs mass satisfying the experimental constraints. 

For large $\tan\beta$, constraints from flavour changing observables 
are very important. For most values of $c_{24}$, $b\to s \gamma$ is 
the most constraining observable. This can be seen from the right panel 
of Figure \ref{fig:mhRgam_c24}. However, for $c_{24}$ around one 
the main constraint comes from $B_s\to\mu^+\mu^-$ because in this 
region the CP-odd Higgs turns out to be light, with mass around 
1 TeV at most.

It was argued in Ref.~\cite{Rgam_g2} that stau-induced enhanced 
$h\to\gamma\gamma$ rate is correlated with SUSY contribution
 to $(g-2)_{\mu}$ and that enhancement of $h\to\gamma\gamma$ rate 
by a few tens of percent typically leads to $(g-2)_{\mu}$ within 
$2\sigma$ from the experimental central value \cite{BNL,Davier_g2SM}.  
However, that statement was based on the assumption of the slepton 
mass universality which is strongly violated in our case by the RG 
effects of the $\tau$ Yukawa coupling. In consequence, we found 
that SUSY contribution to $(g-2)_{\mu}$ is quite small. 
For $c_{24}>2/3$, this contribution is positive, as preferred by 
the experiment, but smaller than approximately $4\times10^{-10}$ 
so about $3\sigma$ below the experimental value. For $c_{24}<2/3$, 
the SUSY contribution to $(g-2)_{\mu}$ (dominated by the chargino-sneutrino 
loop which sign is given by the sign of $\mu M_2$ \cite{Moroi,Stockinger}) 
is negative with the absolute value below about $10^{-9}$. 
Therefore, $(g-2)_{\mu}$ slightly favours $c_{24}>2/3$.

We found that Yukawa-unified solutions with enhanced $\hgamgam$ rate 
have typically large and negative $A$-terms. $A_0/m_{16}$ is between 
-3 and -2, except the region with $c_{24}\in(0.5,0.6)$ where $A_0/m_{16}$ 
between -2.5 and +0.5 is possible. Large negative $A$-terms at the 
GUT scale are generally needed to generate large enough stop mixing 
to account for the observed Higgs boson mass. Nevertheless, we expect 
that large negative values of $A$-terms at the GUT scale are strictly 
related to our assumption of intergenerational degeneracy of the soft 
sfermion masses because it was shown that if at the GUT scale the 
first two-generations sfermions are much heavier than the third-generation 
ones large stop mixing can be generated with small, or even vanishing, 
$A$-terms at the GUT scale \cite{Higgs_IMH}. It is beyond the scope 
of the present work to study in detail the intergenerational splitting 
of scalar masses in the context of top-bottom-tau Yukawa unification 
and we leave it for a future work.

\subsection{Predictions for the MSSM spectrum}

\begin{figure}[t!]
\begin{center}
\includegraphics[width=0.49\textwidth]{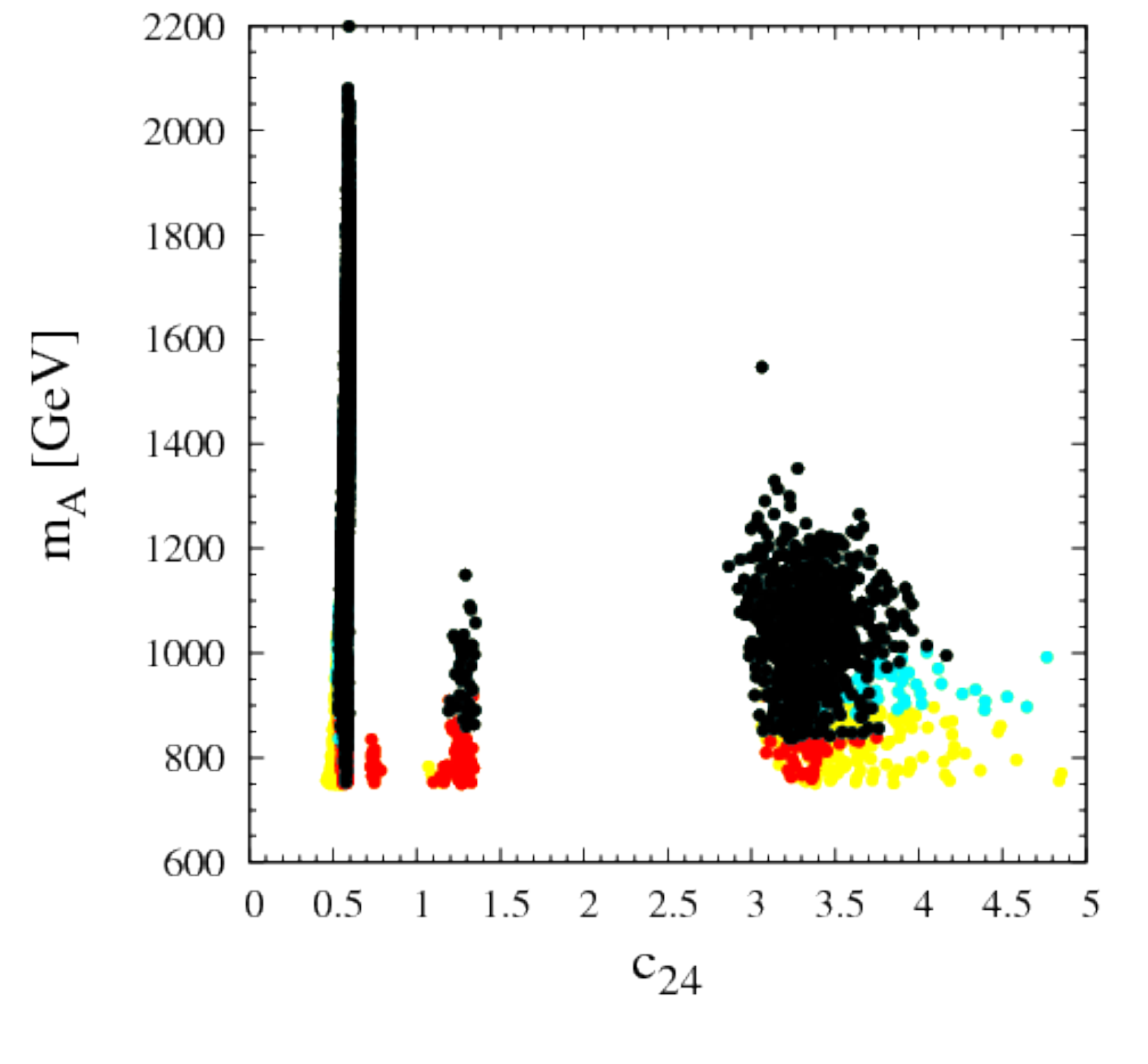}
\hfill
\includegraphics[width=0.49\textwidth]{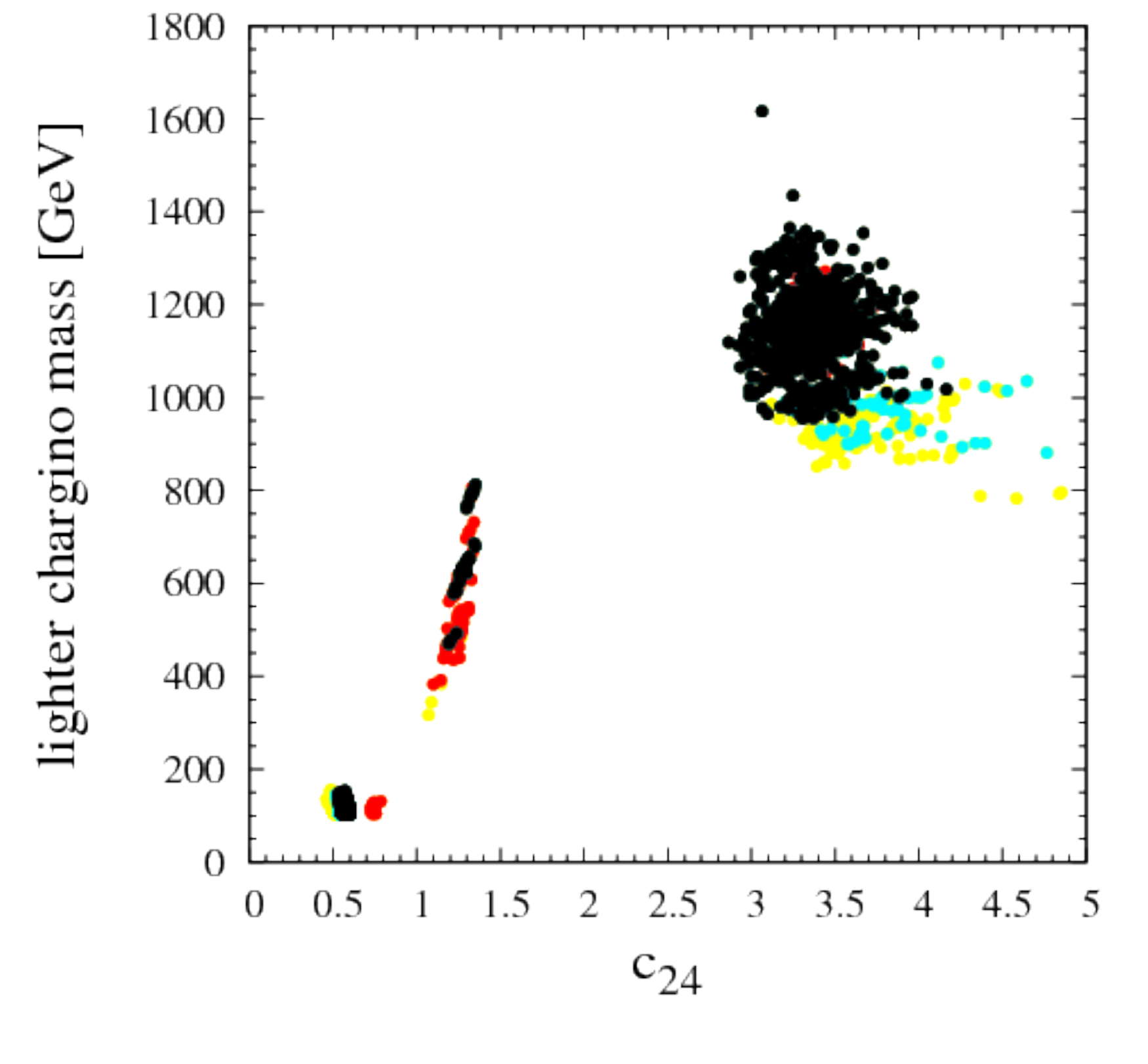}
\caption{Scatter plots of $m_A$ (left panel) and the lighter chargino 
mass (right panel) versus $c_{24}$ for the points with $R<1.1$ 
and $R_{\gamma\gamma}>1.1$. 
The colour coding is the same as in Figure \ref{fig:c1c2}.
} 
\label{fig:mAmcharg_c24}
\end{center}
\end{figure}

The MSSM spectrum is substantially different for different regions 
of $c_{24}$. Examples of such differences, for the pseudoscalar and for 
the lighter chargino masses, are shown in Figure \ref{fig:mAmcharg_c24}.
In the following we discuss separately $c_{24}<1$ 
and $c_{24}>1$, corresponding to the dominant contribution to the gluino 
mass coming from the SUSY breaking $F$-term in the singlet  
and non-singlet representations, respectively.

\subsubsection{Singlet {\boldmath $F$}-term domination}

\begin{figure}[t!]
\begin{center}
\includegraphics[width=0.49\textwidth]{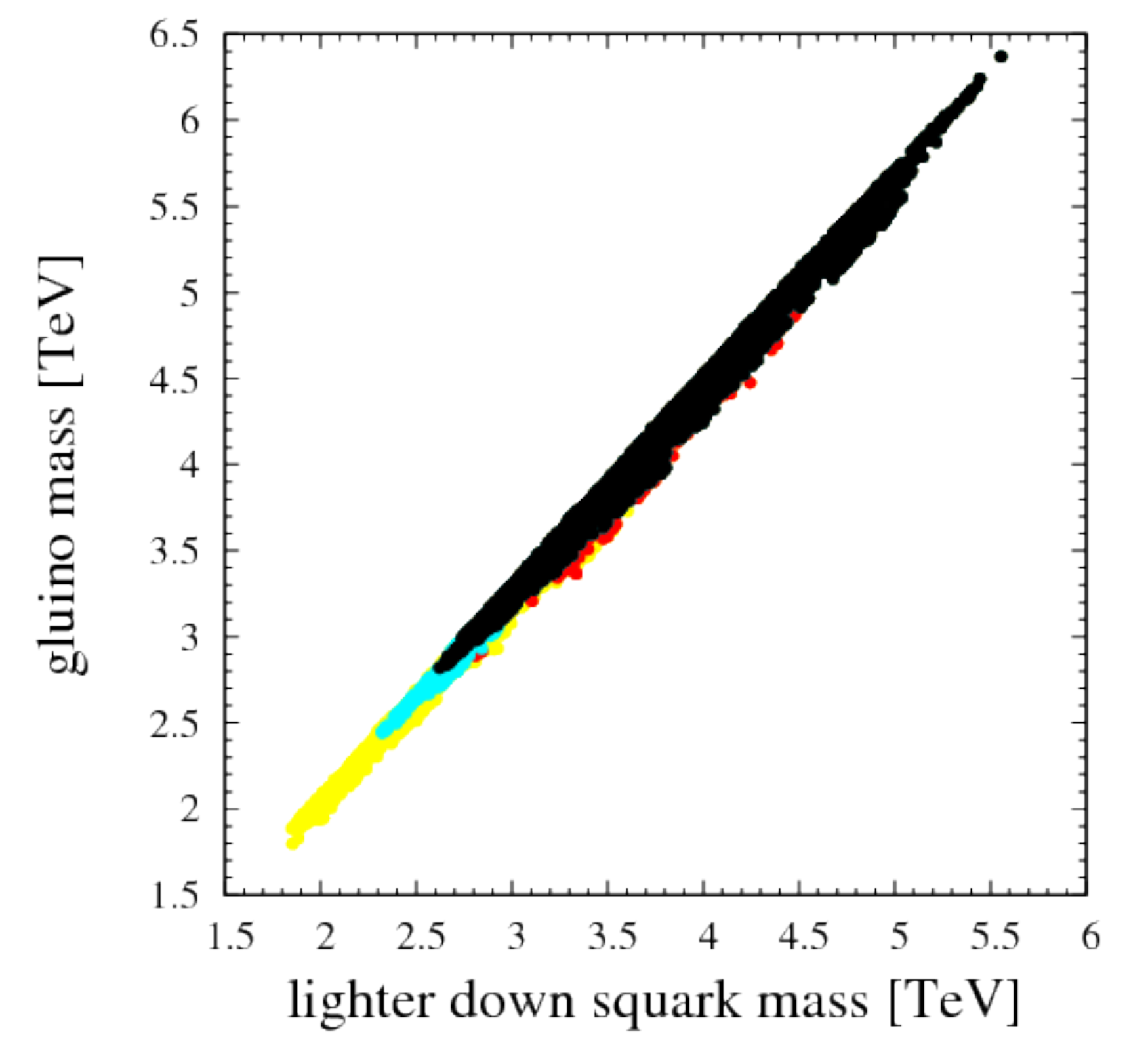}
\hfill
\includegraphics[width=0.49\textwidth]{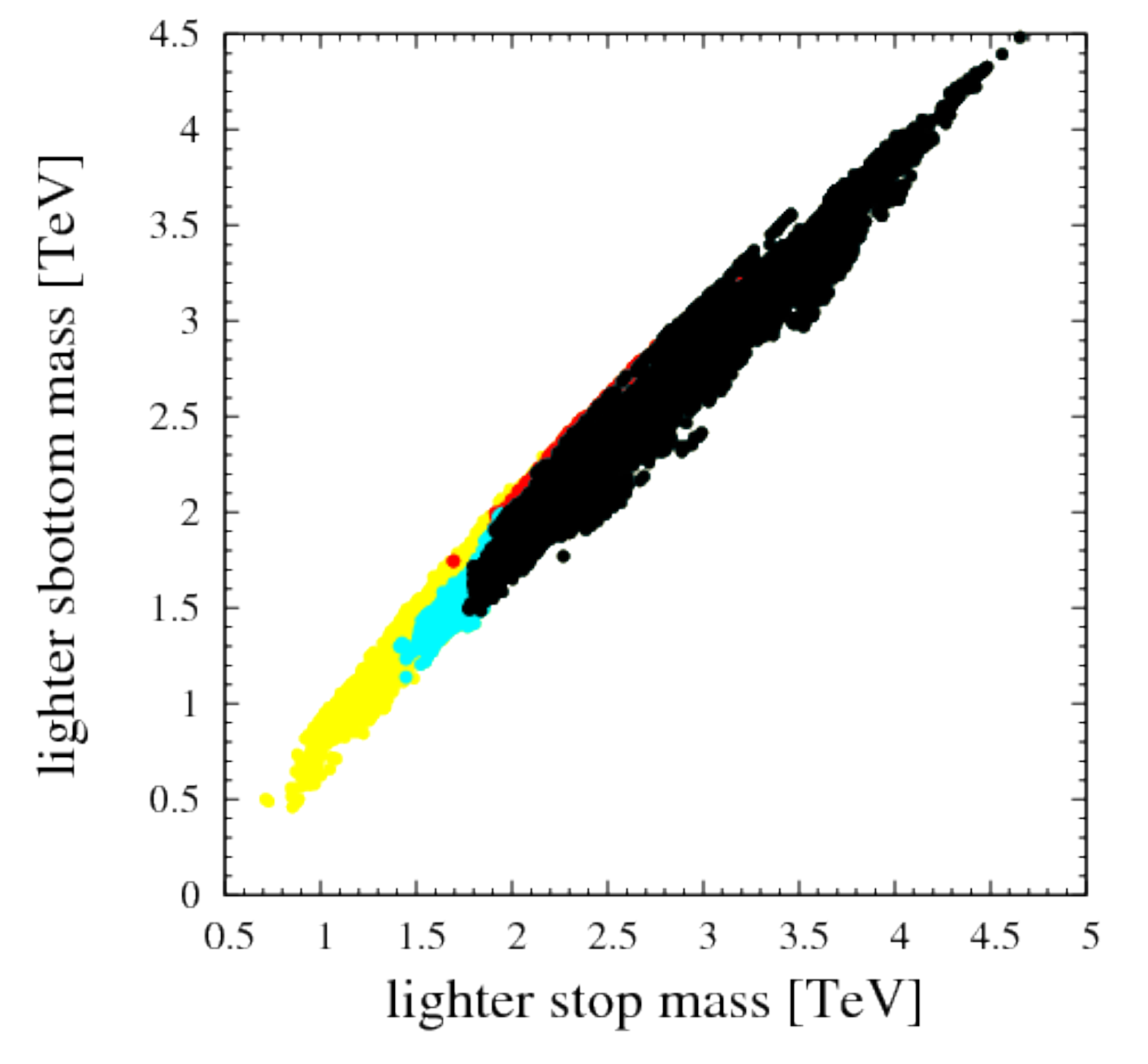}
\caption{Gluino mass versus the right-handed down quark mass (left panel) 
and the lighter sbottom mass versus the lighter stop mass (right panel)  
for the points with $R<1.1$, $R_{\gamma\gamma}>1.1$ and $c_{24}<1$. 
The colour coding is the same as in Figure \ref{fig:c1c2}.
} 
\label{fig:masses_c24large}
\end{center}
\end{figure}

For $c_{24}<1$, the dominant contribution to the gluino mass comes for 
the singlet $F$-term. However, enhanced $h\to\gamma\gamma$ rate requires 
non-negligible contribution of the non-singlet $F$-term. There are two 
separate regions with $c_{24}\in(0.5,0.6)$ and $c_{24}\in(0.7,0.8)$. 
Partial cancellation between the contributions to $M_2$ from the singlet and 
non-singlet $F$-terms results in a wino-like LSP with mass above 100 GeV, 
due to the LEP constraint on the chargino mass which is almost 
degenerate with the LSP. In consequence, also the lightest stau mass 
is above 100 GeV. Nevertheless, significant enhancement of the 
$h\to\gamma\gamma$ rate, especially for $c_{24}\in(0.5,0.6)$, 
is possible. Since the LSP in this region is wino-like its thermal 
relic abundance is much too small to explain the observed relic 
abundance of dark matter \cite{welltempered} unless non-standard 
cosmological history is assumed.

Even though the wino-like chargino is very light in this scenario it is 
very challenging to discover it at the LHC because it decays predominantly 
to the LSP and a very soft pion \cite{winoLSP_charginosearch}. It is also 
rather difficult, but not impossible, to probe this scenario with the LHC 
searches for coloured sparticles because the gluino mass is above 2.8 TeV 
while the right-handed down squark, which is the lightest first-generation 
squark, have mass at least 2.6 TeV, as can be seen from the left panel 
of Figure \ref{fig:masses_c24large}. Such SUSY states could be 
discovered at the 14 TeV LHC but only if their masses are close to the 
lower bounds quoted above \cite{LHCreach}. The third-generation 
squarks are lighter than those from the first two generations. 
Nevertheless, the mass of the lightest sbottom (stop) is above about 1.5 (1.8) 
TeV which will require very large statistics to discover it at the LHC.

It is worth pointing out that constraints from $b\to s \gamma$ should 
not be used for an ultimate exclusion of a given MSSM model. This is 
because the MSSM prediction for BR$(b\to s \gamma)$ can be easily 
affected by a flavour violating gluino contribution, see 
e.g.~\cite{Chankowski,Okumura,YU_Hall}. It was recently argued in 
\cite{YU_Hall} that even if BR$(b\to s \gamma)$ calculated assuming 
minimal flavour violation disagrees with the experimental result the 
flavour violating gluino contribution can  brought it in agreement 
with the experimental data without large fine-tuning. Admitting such 
additional contributions the blue points in our figures should be 
considered allowed. In such a case the lower bound for the gluino 
(the right-handed down squark) mass is reduced to 2.4 (2.3) TeV so 
should be probed with ${\mathcal O}(100\, {\rm fb}^{-1})$ at the 
14 TeV LHC. The third-generation squarks can also be lighter in 
such a case. The lower bound on the lightest sbottom (stop) 
reduces to 1.1 (1.4) TeV making them accessible in the early stage of 
the LHC run after the energy upgrade.

The most promising signature of the model at the LHC is a light 
pseudoscalar Higgs which can be arbitrarily close to the present 
experimental lower bound on $m_A$. Interestingly, one branch of 
solutions with $c_{24}\in(0.7,0.8)$ have also quite strong upper 
bound on $m_A$ of about 900 GeV which will be entirely probed in 
the very early stage of the 14 TeV LHC operation. However, this branch of solutions is incompatible with the recent measurement of BR$(B_s\to\mu^+\mu^-)$.

\subsubsection{Non-singlet {\boldmath $F$}-term domination}

In the region of the parameter space with $c_{24}>1$, corresponding 
to the gluino mass generated mainly by the non-singlet $F$-term, 
the LSP is bino-like which, in contrast to the wino-like LSP, 
can be below 100 GeV without violating the collider constraints. 
Therefore, in this case the lightest stau can also be below 100 GeV 
and the lower limit on its mass from LEP is about 82-90 GeV 
(depending on the stau mixing angle and the mass difference with the LSP). 
That is why the $h\to\gamma\gamma$ rate can be somewhat larger than for 
the $c_{24}<1$.

The lightest chargino is also wino-like when  $c_{24}>1$ but it is 
significantly heavier than in the $c_{24}<1$ case, as seen from 
the right panel of 
Figure \ref{fig:mAmcharg_c24}. For $c_{24}\in(3,4)$ it is in the 
$1\div1.5$ TeV 
range, while for $c_{24}\in(1.2,1.4)$ it is lighter (due to larger 
cancellation between the singlet and non-singlet $F$-term contributions), 
between about 400 and 800 GeV. The latter region could be, in principle, 
interesting from the point of view of the LHC phenomenology. This is 
because the lower mass limits for wino decaying via an on-shell stau, 
which is the dominant decay channel in our case, reach about 300 GeV in 
certain circumstances \cite{ATLAS_wino_tau,CMS_wino}. However, if one 
demands that the upper bound on $\Omega_{\rm LSP}$ is satisfied than the 
mass splitting between the stau and the LSP is below about 10 GeV so taus 
produced by decays of intermediate staus are soft and very hard to detect. 
In consequence, for mass splitting allowing efficient stau co-annihilation  
the LHC does not provide any constraints on the wino mass.

Searches for staus are even more challenging because the production 
cross-section is much smaller than the wino production cross-section 
and taus resulting from stau decays are also very soft.

\begin{figure}[t!]
\begin{center}
\includegraphics[width=0.49\textwidth]{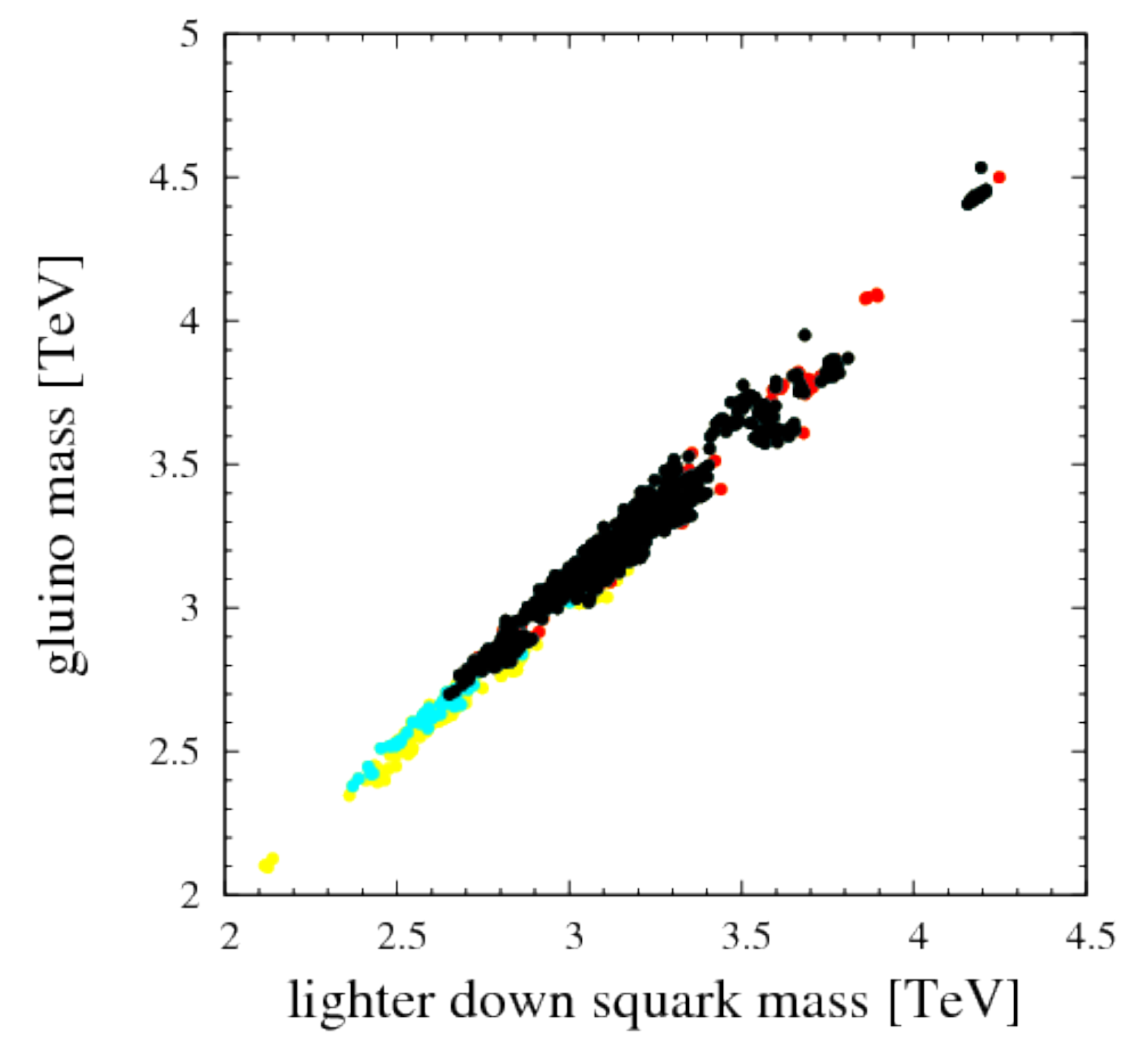}
\hfill
\includegraphics[width=0.49\textwidth]{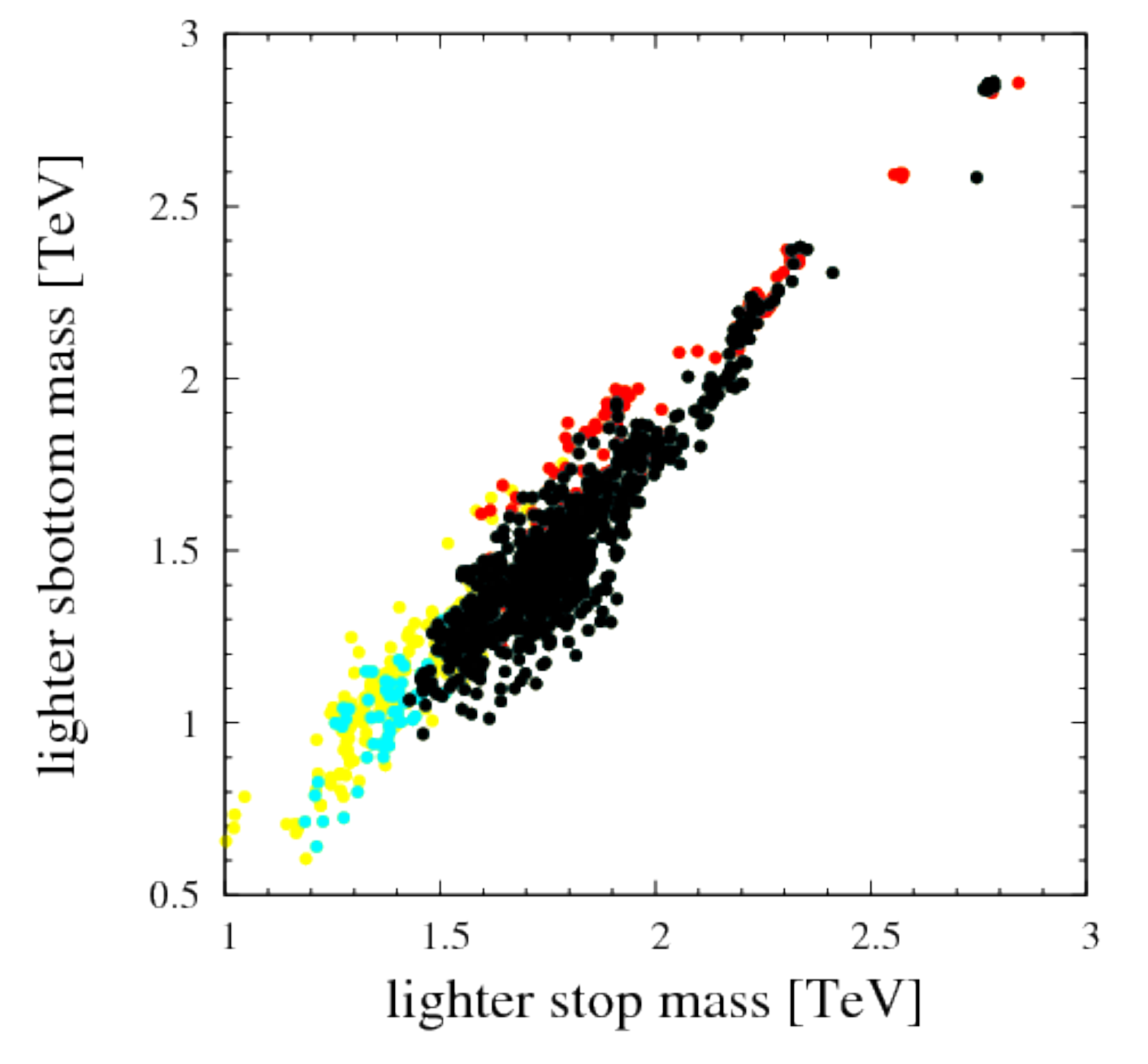}
\caption{The same as in Figure \ref{fig:masses_c24large} but for $c_{24}>1$.
} 
\label{fig:masses_c24small}
\end{center}
\end{figure}

The squarks of the first two generations and the gluino in the 
$c_{24}>1$ case are also rather heavy. The gluino mass and the right-handed 
down squark are heavier than about 2.7 TeV, as seen from the left panel of 
Figure \ref{fig:masses_c24small}. However, the third-generation squarks 
are somewhat lighter.  The lighter stop mass is above 1.4 TeV. 
The (mostly right-handed) sbottom can be as light as 1 TeV so not far 
away from the present experimental lower limit of about 600 GeV 
\cite{Atlas_sbottom}.

Relaxing the constraint from $b\to s \gamma$ reduces the lower bound 
on the gluino and right-handed down squark masses to 2.4 TeV. 
The lower limit on the stop mass is reduced in such a case to 1.2 TeV. 
What is the most interesting, the lower limit on the sbottom mass 
becomes about 600 GeV so the sbottom can be around the corner.

A very interesting prediction of the model with the non-singlet 
$F$-term domination  is that the CP-odd Higgs is relatively light. 
It can be seen from the left panel of Figure \ref{fig:mAmcharg_c24} 
that $m_A$ is below about 1.6 TeV.
It was recently shown in Ref.~\cite{mA_prospects} that majority of pMSSM points with $\tan\beta\sim45$, which is a typical value for points with top-bottom-tau Yukawa unification, and $m_A$ about 1.5 TeV can be excluded with 150 fb$^{-1}$ of data at the 14 TeV LHC.\footnote{We would like to thank Ian Lewis for turning our attention to Ref.~\cite{mA_prospects}.}
Points that could avoid exclusion are characterized by very large SUSY threshold correction to the bottom mass or by substantial branching ratio of the heavy Higgses to SUSY particles. The former case does not apply to our model since the condition of Yukawa unification implies that the SUSY threshold correction to the bottom mass can not be large (see Subsection \ref{subsec:YU}). We also checked with SUSYHIT \cite{Susyhit} that in this scenario the branching ratio of $H$ and $A$ to stau pairs is always below ten percent ($H$/$A$ decays to neutralinos are completely negligible). 
In addition, the 14 TeV LHC is expected to deliver much more luminosity than 150 fb$^{-1}$ (used in the study of Ref.~\cite{mA_prospects}), see e.g. Ref.~\cite{LHC_lumi}. 
Therefore, it is likely that the whole region  of parameter space with the non-singlet 
$F$-term domination (or at least large part of it) can be ruled out by the heavy MSSM Higgs searches at the 
14 TeV LHC.

 It should be emphasized that the lightness of the CP-odd 
Higgs is tightly connected with the assumption of the enhanced 
$h\to\gamma\gamma$ rate. This is confirmed by the analysis performed 
in Ref.~\cite{YUreview} where the case when only {\bf 24} $F$-term 
contributes to the gaugino masses (which corresponds to the limiting 
case $c_{24}\to\infty$ of the present model) was considered 
and enhancement of the $h\to\gamma\gamma$ rate was not required. 
It was shown that 
in such a case there is no sharp prediction for $m_A$ which could 
be in the multi-TeV range, well outside of the LHC reach.

A qualitative argument for the prediction of small $m_A$ is following. For large $\tan\beta$ and after imposing REWSB $m_A^2$ is well approximated by $m_{H_d}^2-m_{H_u}^2$. RG contribution to $m_{H_d}^2-m_{H_u}^2$ from gauginos is $\mathcal{O}(0.1)M_3^2$, while the contributions from a universal soft scalar masses and soft trilinear terms are negative, see e.g.\ Ref.~\cite{YUreview}. This imply $m_A\lesssim 0.1m_{\tilde{g}}$ with the inequality saturated in the limit $M_3\gg m_0,A_0$. Too light pseudoscalar Higgs can be avoided with the help of the $D$-term contribution, cf. eq.~(\ref{scalarDterm}), since $D>0$ gives positive contribution to $m_{H_d}^2-m_{H_u}^2$. 
However, the requirement of light staus to enhance $\hgamgam$ rate and rather heavy stops (to account for the observed Higgs mass) implies that the positive RGE contribution from gauginos to the stop masses should be large as compared to $m_{16}$ and the negative RGE contribution proportional to the Yukawa couplings. This happens only if  $m_{16}/M_3$ is not too large. Since $D/m_{16}^2<1/3$, in order to have positive soft mass squared at the GUT scale, the $D$-term contribution to $m_{H_d}^2-m_{H_u}^2$, hence also to $m_A^2$,  is also constrained from above by the requirement of large splitting between the stau and stop masses.

\section{Conclusions}
\label{sec:concl}

We studied enhanced $\hgamgam$ rate induced by light staus with a 
strong left-right mixing. We found that the requirement of 
a substantial enhancement of the $\hgamgam$ rate leads to 
strong constraints on the gaugino masses at the GUT scale: 
$|M_1/M_3|\lesssim1/2$ and/or $|M_2/M_3|\lesssim1/4$. 
This constraint follows from the requirement of a neutral LSP, 
the LHC limits on the gluino mass and the one-loop RGE prediction 
for the low-energy gaugino masses. Therefore, it is applicable 
not only in the (R-parity conserving) MSSM but also in many of 
its extensions such as the NMSSM.

We made a successful attempt to accommodate the MSSM spectrum 
with strongly-mixed staus inducing enhanced $\hgamgam$ rate 
in SO(10) models predicting top-bottom-tau Yukawa unification.    
We argued that a substantial enhancement of the $\hgamgam$ rate 
is possible only for the negative sign of $\mu$, so future measurements 
of the $\hgamgam$ rate may discriminate between models with 
different signs of $\mu$.
Assuming the $D$-term splitting of scalar masses, we identified 
patterns of the gaugino masses that allow for top-bottom-tau Yukawa 
unification and enhanced $\hgamgam$ rate. These patterns can be 
accommodated in well-motivated models of SUSY breaking such as 
mirage mediation or gravity mediation with the SUSY breaking 
$F$-term that is a mixture of the singlet and non-singlet 
representations of SO(10).

We investigated in detail a particular scenario in which the 
gaugino masses are generated by a combination of the singlet $F$-term 
and the $F$-term in ${\bf 24}\subset{\bf 54}$ of $SU(5)\subset SO(10)$. 
We found that the $\hgamgam$ rate in this scenario can be enhanced 
by more than 30\% in agreement with the phenomenological constraints, 
including vacuum metastability bounds. In order to account for 
enhanced $\hgamgam$ rate and top-bottom-tau Yukawa unification 
the singlet and non-singlet $F$-term contributions to the gluino mass 
should be of the same order. Nevertheless, these contributions can 
differ by a factor of a few and only some ratios of these contributions 
can be consistent with the experimental constraints.

There are some phenomenological differences between models depending 
on whether the gluino mass arises mainly from the singlet or the 
non-singlet $F$-term. If the singlet $F$-term dominates, the LSP is 
wino-like with too small thermal relic abundance to account for the 
observed energy density of dark matter. On the other hand, if the 
non-singlet $F$-term dominates the LSP is bino-like and its thermal 
relic abundance can be brought to cosmologically acceptable 
values due to efficient coannihilation with staus.

In any case the resulting spectrum of coloured sparticles is rather heavy. 
The lower bounds on most of them are only slightly below 3 TeV so it 
will require a lot of data at the 14 TeV LHC to start to probe them. 
The exception is the lightest sbottom which in the non-singlet $F$-term 
domination case may have mass around 1 TeV. The best prospects for 
testing the model is due to the pseudoscalar Higgs which generically 
has mass close to the present experimental lower bound.     

It is interesting to note that in spite of the correlation between the soft masses for squarks and sleptons at the GUT scale dictated by the structure of SO(10) GUTs, in the low-energy spectrum very light stau, with mass ${\mathcal O}(100\ \GeV)$ to account for enhanced $\hgamgam$ rate, and heavy squarks can be simultaneously present. This is possible due to RG effect of gluino which contribute substantially to squark masses without affecting the slepton masses. The mass splitting between staus and the first two generation squarks is additionally enhanced due to large tau Yukawa coupling which reduces the stau masses via RGEs.

We also found that the $b\to s\gamma$ and $B_s\to \mu^+\mu^-$ 
constraints push the supersymmetric spectrum up in a significant way. 
If these are neglected, due to non-minimally-flavour-violating contributions, 
most of the coloured sparticles could be accessible in the early 
stage of the next LHC run.

The only observable that cannot be fitted in this model is $(g-2)_{\mu}$
since slepton masses of the second generation are much heavier than staus because only the latter acquire a negative RGE contribution proportional to large Yukawa coupling.  
In principle, the $(g-2)_{\mu}$ anomaly could 
be explained in SO(10) models in the framework of the NMSSM since 
at large $\tan\beta$ additional contributions to the Higgs mass 
due to mixing with the singlet \cite{NMSSMmixing} allow to lower 
the scale of superpartner masses. 
We plan to investigate this issue in the future.

\section*{Acknowledgments}

We would like to thank Artur Kalinowski, Ian Lewis, Stuart Raby, Kazuki Sakurai and James Wells 
for discussions.
This work is a part of the ``Implications of the Higgs boson discovery on supersymmetric extensions of the Standard Model'' project funded within the HOMING PLUS programme of the Foundation for Polish Science. 
This work has also been partially supported by the European Commission under the contract PITN-GA-2009-237920 UNILHC. 
MO and SP have been supported by National Science Centre under research 
grants DEC-2011/01/M/ST2/02466, DEC-2012/04/A/ST2/00099, 
DEC-2012/05/B/ST2/02597.
MB has been partially supported by the MNiSW grant IP2012 030272
and the Foundation for Polish Science through its programme START. 
MO would like to thank 
the CERN theory division for hospitality during the final stage of 
this work. 
MB thanks the Galileo Galilei Institute for Theoretical Physics
and INFN for hospitality and partial support during the final stage of this work.



\begin{thebibliography}{99}

\bibitem{Atlas_discovery}
  G.~Aad {\it et al.}  [ATLAS Collaboration],
  Phys.\ Lett.\ B {\bf 716} (2012) 1
  [arXiv:1207.7214 [hep-ex]].

\bibitem{CMS_discovery}
  S.~Chatrchyan {\it et al.}  [CMS Collaboration],
  Phys.\ Lett.\ B {\bf 716} (2012) 30
  [arXiv:1207.7235 [hep-ex]].

\bibitem{Atlas_gamma}
  ATLAS Collaboration,
  ATLAS-CONF-2013-012.

\bibitem{Atlas_WWZZ}
  ATLAS Collaboration,
ATLAS-CONF-2013-013, ATLAS-CONF-2013-030.

\bibitem{CMS_gamma}
CMS Collaboration, CMS-PAS-HIG-13-001.

\bibitem{CMS_WWZZ}
CMS Collaboration,  CMS-PAS-HIG-13-002, CMS-PAS-HIG-13-003.

\bibitem{CMS_gamma_old}
CMS Collaboration, CMS-PAS-HIG-12-015.

\bibitem{DjouadiMSSMreview}
  A.~Djouadi,
  Phys.\ Rept.\  {\bf 459} (2008) 1
  [hep-ph/0503173].

\bibitem{Carena_stau}
  M.~Carena, S.~Gori, N.~R.~Shah and C.~E.~M.~Wagner,
  JHEP {\bf 1203} (2012) 014
  [arXiv:1112.3336 [hep-ph]];
  M.~Carena, S.~Gori, N.~R.~Shah, C.~E.~M.~Wagner and L.~-T.~Wang,
  JHEP {\bf 1207} (2012) 175
  [arXiv:1205.5842 [hep-ph]].

\bibitem{Ellwanger_reducedhbb}
  U.~Ellwanger,
  JHEP {\bf 1203} (2012) 044
  [arXiv:1112.3548 [hep-ph]].

\bibitem{Cao}
  J.~-J.~Cao, Z.~-X.~Heng, J.~M.~Yang, Y.~-M.~Zhang and J.~-Y.~Zhu,
  JHEP {\bf 1203} (2012) 086
  [arXiv:1202.5821 [hep-ph]].

\bibitem{Carena_diphoton}
  M.~Carena, I.~Low and C.~E.~M.~Wagner,
  JHEP {\bf 1208} (2012) 060
  [arXiv:1206.1082 [hep-ph]].

\bibitem{diphoton_scalar}
  A.~Delgado, G.~Nardini and M.~Quiros,
  Phys.\ Rev.\ D {\bf 86} (2012) 115010
  [arXiv:1207.6596 [hep-ph]];
  M.~A.~Ajaib, I.~Gogoladze and Q.~Shafi,
  Phys.\ Rev.\ D {\bf 86} (2012) 095028
  [arXiv:1207.7068];
 M.~Chala,
  JHEP {\bf 1301} (2013) 122
  [arXiv:1210.6208 [hep-ph]];
  B.~Swiezewska and M.~Krawczyk,
  Phys.\  Rev.\  D {\bf 88} (2013) 035019
  [arXiv:1212.4100 [hep-ph]].

\bibitem{diphoton_boson}
  D.~Carmi, A.~Falkowski, E.~Kuflik and T.~Volansky,
  JHEP {\bf 1207} (2012) 136
  [arXiv:1202.3144 [hep-ph]];
B.~Bellazzini, C.~Csaki, J.~Hubisz, J.~Serra and J.~Terning,
  JHEP {\bf 1211} (2012) 003
  [arXiv:1205.4032 [hep-ph]];
  A.~Alves, E.~Ramirez Barreto, A.~G.~Dias, C.~A.~de S.Pires, F.~S.~Queiroz and P.~S.~Rodrigues da Silva,
  Eur.\ Phys.\ J.\ C {\bf 73} (2013) 2288
  [arXiv:1207.3699 [hep-ph]];
  T.~Abe, N.~Chen and H.~-J.~He,
  JHEP {\bf 1301} (2013) 082
  [arXiv:1207.4103 [hep-ph]].

\bibitem{diphoton_fermion}
 H.~An, T.~Liu and L.~-T.~Wang,
  Phys.\ Rev.\ D {\bf 86} (2012) 075030
  [arXiv:1207.2473 [hep-ph]];
 A.~Joglekar, P.~Schwaller and C.~E.~M.~Wagner,
  JHEP {\bf 1212} (2012) 064
  [arXiv:1207.4235 [hep-ph]];
  N.~Arkani-Hamed, K.~Blum, R.~T.~D'Agnolo and J.~Fan,
  JHEP {\bf 1301} (2013) 149
  [arXiv:1207.4482 [hep-ph]];
 N.~Bonne and G.~Moreau,
  Phys.\ Lett.\ B {\bf 717} (2012) 409
  [arXiv:1206.3360 [hep-ph]];
  A.~Carmona and F.~Goertz,
  arXiv:1301.5856 [hep-ph].


\bibitem{gamma_chargino}
  J.~A.~Casas, J.~M.~Moreno, K.~Rolbiecki and B.~Zaldivar,
  arXiv:1305.3274 [hep-ph].

\bibitem{Staub_stau}
  L.~Basso and F.~Staub,
  Phys.\ Rev.\ D {\bf 87} (2013) 015011
  [arXiv:1210.7946 [hep-ph]].

\bibitem{mosp_tbuni}
  M.~Olechowski and S.~Pokorski,
  Phys.\ Lett.\ B\ {\bf 214} (1988) 393.

\bibitem{YU_tbtau}
  B.~Ananthanarayan, G.~Lazarides and Q.~Shafi,
  Phys.\ Rev.\ D {\bf 44} (1991) 1613; 
  G.~W.~Anderson, S.~Raby, S.~Dimopoulos and L.~J.~Hall,
  Phys.\ Rev.\ D {\bf 47} (1993) 3702
  [hep-ph/9209250]. \\
For some more recent works see e.g. \\
   D.~Guadagnoli, S.~Raby and D.~M.~Straub,
  JHEP {\bf 0910} (2009) 059
  [arXiv:0907.4709 [hep-ph]].
I.~Gogoladze, R.~Khalid, S.~Raza and Q.~Shafi,
  JHEP {\bf 1012} (2010) 055
  [arXiv:1008.2765 [hep-ph]];
  I.~Gogoladze, Q.~Shafi and C.~S.~Un,
  Phys.\ Lett.\ B {\bf 704} (2011) 201
  [arXiv:1107.1228 [hep-ph]];
  A.~S.~Joshipura and K.~M.~Patel,
  Phys.\ Rev.\ D {\bf 86} (2012) 035019
  [arXiv:1206.3910 [hep-ph]];
  M.~Adeel Ajaib, I.~Gogoladze, Q.~Shafi and C.~S.~Un,
  JHEP {\bf 1307} (2013) 139
  [arXiv:1303.6964 [hep-ph]].

\bibitem{bop_F24}
  M.~Badziak, M.~Olechowski and S.~Pokorski,
  JHEP\ {\bf 1108} (2011) 147
  [arXiv:1107.2764 [hep-ph]].

\bibitem{mbks}
  M.~Badziak and K.~Sakurai,
  JHEP {\bf 1202} (2012) 125
  [arXiv:1112.4796 [hep-ph]].

\bibitem{Hall}
  L.~J.~Hall, R.~Rattazzi, U.~Sarid,
  Phys.\ Rev.\  {\bf D50 } (1994)  7048-7065.
  [hep-ph/9306309].

\bibitem{Carena}
  M.~S.~Carena, M.~Olechowski, S.~Pokorski, C.~E.~M.~Wagner,
  Nucl.\ Phys.\  {\bf B426 } (1994)  269-300.
  [hep-ph/9402253].

\bibitem{gaugeuni}
  M.~S.~Carena, S.~Pokorski and C.~E.~M.~Wagner,
  Nucl.\ Phys.\ B {\bf 406} (1993) 59
  [hep-ph/9303202].


\bibitem{Olechowski}
  M.~Olechowski, S.~Pokorski,
  Phys.\ Lett.\  {\bf B344 } (1995)  201-210.
  [hep-ph/9407404].

\bibitem{YU_125Higgs}
  I.~Gogoladze, Q.~Shafi and C.~S.~Un,
  JHEP {\bf 1207} (2012) 055
  [arXiv:1203.6082 [hep-ph]];
  I.~Gogoladze, Q.~Shafi and C.~S.~Un,
  JHEP {\bf 1208} (2012) 028
  [arXiv:1112.2206 [hep-ph]].

\bibitem{YUreview}
  M.~Badziak,
  Mod.\ Phys.\ Lett.\ A {\bf 27} (2012) 1230020
  [arXiv:1205.6232 [hep-ph]].

\bibitem{Dterm}
  Y.~Kawamura, H.~Murayama, M.~Yamaguchi,
  Phys.\ Rev.\  {\bf D51 } (1995)  1337-1352.
  [hep-ph/9406245].

\bibitem{Murayama}
  H.~Murayama, M.~Olechowski, S.~Pokorski,
  Phys.\ Lett.\  {\bf B371 } (1996)  57-64.
  [hep-ph/9510327].

\bibitem{mirage1}
  K.~Choi, A.~Falkowski, H.~P.~Nilles and M.~Olechowski,
  Nucl.\ Phys.\ B {\bf 718} (2005) 113
  [hep-th/0503216].

\bibitem{mirage2}
  K.~Choi, K.~S.~Jeong and K.~-i.~Okumura,
  JHEP {\bf 0509} (2005) 039
  [hep-ph/0504037].

\bibitem{mirage3}
  A.~Falkowski, O.~Lebedev and Y.~Mambrini,
  JHEP {\bf 0511} (2005) 034
  [hep-ph/0507110].

\bibitem{mirage4}
  O.~Lebedev, H.~P.~Nilles and M.~Ratz,
  Phys.\ Lett.\ B {\bf 636} (2006) 126
  [hep-th/0603047].

\bibitem{mirage5}
  K.~Choi and H.~P.~Nilles,
  JHEP {\bf 0704} (2007) 006
  [hep-ph/0702146 [HEP-PH]].

\bibitem{mirage6}
  V.~Lowen and H.~P.~Nilles,
  Phys.\ Rev.\ D {\bf 77} (2008) 106007
  [arXiv:0802.1137 [hep-ph]].

\bibitem{mirage7}
  M.~Badziak, S.~Krippendorf, H.~P.~Nilles and M.~W.~Winkler,
  JHEP {\bf 1303} (2013) 094
  [arXiv:1212.0854 [hep-ph]].


\bibitem{DjouadiSMreview}
  A.~Djouadi,
  Phys.\ Rept.\  {\bf 457} (2008) 1
  [hep-ph/0503172].

\bibitem{Djouadi_reducedcross}
  A.~Djouadi,
  Phys.\ Lett.\ B {\bf 435} (1998) 101
  [hep-ph/9806315].

\bibitem{Rgam_g2}
  G.~F.~Giudice, P.~Paradisi, A.~Strumia,
  JHEP {\bf 1210} (2012) 186
  [arXiv:1207.6393 [hep-ph]].

\bibitem{AkulaNath}
    S.~Akula and P.~Nath,
  Phys.\  Rev.\  D 87, {\bf 115022} (2013)
  [arXiv:1304.5526 [hep-ph]].

\bibitem{LEPstau}
  J.~Beringer {\it et al.}  [Particle Data Group Collaboration],
  Phys.\ Rev.\ D {\bf 86} (2012) 010001;
  A.~Heister {\it et al.}  [ALEPH Collaboration],
  Phys.\ Lett.\ B {\bf 526} (2002) 206
  [hep-ex/0112011];
  J.~Abdallah {\it et al.}  [DELPHI Collaboration],
  Eur.\ Phys.\ J.\ C {\bf 31} (2003) 421
  [hep-ex/0311019];
  P.~Achard {\it et al.}  [L3 Collaboration],
  Phys.\ Lett.\ B {\bf 580} (2004) 37
  [hep-ex/0310007];
  G.~Abbiendi {\it et al.}  [OPAL Collaboration],
  Eur.\ Phys.\ J.\ C {\bf 32} (2004) 453
  [hep-ex/0309014]; 
LEP2 SUSY working group, \url{http://lepsusy.web.cern.ch/lepsusy/www/sleptons_summer04/slep_final.html}.

\bibitem{stability_Rattazzi}
  R.~Rattazzi and U.~Sarid,
  Nucl.\ Phys.\ B {\bf 501} (1997) 297
  [hep-ph/9612464].

\bibitem{stability_Hisano}
  J.~Hisano and S.~Sugiyama,
  Phys.\ Lett.\ B {\bf 696} (2011) 92
   [Erratum-ibid.\ B {\bf 719} (2013) 472]
  [arXiv:1011.0260 [hep-ph]].

\bibitem{stability_Carena}
  M.~Carena, S.~Gori, I.~Low, N.~R.~Shah and C.~E.~M.~Wagner,
  JHEP {\bf 1302} (2013) 114
  [arXiv:1211.6136 [hep-ph]].

\bibitem{stability_Kitahara}
  T.~Kitahara and T.~Yoshinaga,
  JHEP {\bf 1305} (2013) 035
  [arXiv:1303.0461 [hep-ph]].

\bibitem{BaerDM}
  H.~Baer, S.~Kraml, S.~Sekmen, H.~Summy,
  JHEP {\bf 0803 } (2008)  056.
  [arXiv:0801.1831 [hep-ph]].

\bibitem{compressed_Dreiner}
  H.~K.~Dreiner, M.~Kramer and J.~Tattersall,
  Europhys.\ Lett.\  {\bf 99} (2012) 61001
  [arXiv:1207.1613 [hep-ph]].

\bibitem{ATLAS_gluino_bb}
ATLAS Collaboration, ATLAS-CONF-2012-145

\bibitem{CMS_gluino_bb}
CMS Collaboration, CMS-PAS-SUS-12-024

\bibitem{Pierce}
  D.~M.~Pierce, J.~A.~Bagger, K.~T.~Matchev, R.~-j.~Zhang,
  Nucl.\ Phys.\  {\bf B491 } (1997)  3-67.
  [hep-ph/9606211].

\bibitem{Wells_yuk}
  K.~Tobe, J.~D.~Wells,
  Nucl.\ Phys.\  {\bf B663 } (2003)  123-140.
  [hep-ph/0301015].

\bibitem{Blazek}
  T.~Blazek, R.~Dermisek, S.~Raby,
  Phys.\ Rev.\  {\bf D65 } (2002)  115004.
  [hep-ph/0201081].

\bibitem{Baer_heaviergluino}
  H.~Baer, S.~Raza and Q.~Shafi,
  Phys.\ Lett.\ B {\bf 712} (2012) 250
  [arXiv:1201.5668 [hep-ph]].

\bibitem{YU_Rabynew}
  A.~Anandakrishnan, S.~Raby and A.~Wingerter,
  Phys.\ Rev.\ D {\bf 87} (2013) 055005
  [arXiv:1212.0542 [hep-ph]].

\bibitem{metropolis1}
N. Metropolis, A.W. Rosenbluth, M.N. Rosenbluth, A.H. Teller, and E. Teller, 
Journal of Chemical Physics, 21(6):1087-1092, 1953.
\bibitem{metropolis2}
W.K. Hastings, 
Biometrika, 57(1):97--109, 1970.

\bibitem{softsusy}
  B.~C.~Allanach,
  Comput.\ Phys.\ Commun.\  {\bf 143 } (2002)  305-331.
  [hep-ph/0104145].

\bibitem{Micromega}
  G.~Belanger, F.~Boudjema, A.~Pukhov, A.~Semenov,
  Comput.\ Phys.\ Commun.\  {\bf 176 } (2007)  367-382.
  [hep-ph/0607059].

\bibitem{bsg}
  D.~Asner {\it et al.}  [Heavy Flavor Averaging Group],
  arXiv:1010.1589 [hep-ex].

\bibitem{bsgth}
  M.~Misiak, H.~M.~Asatrian, K.~Bieri, M.~Czakon, A.~Czarnecki, T.~Ewerth, A.~Ferroglia, P.~Gambino {\it et al.},
  Phys.\ Rev.\ Lett.\  {\bf 98 } (2007)  022002.
  [hep-ph/0609232].

\bibitem{bsgth2}
 G.~Degrassi, P.~Gambino and G.~F.~Giudice,
  JHEP {\bf 0012} (2000) 009
  [arXiv:hep-ph/0009337].

\bibitem{Bsmumu_comb}
CMS and LHCb collaborations, CMS-PAS-BPH-13-007, LHCb-CONF-2013-012, \url{http://cds.cern.ch/record/1564324}.



\bibitem{Bsmumu_th}
  A.~J.~Buras, J.~Girrbach, D.~Guadagnoli and G.~Isidori,
  Eur.\ Phys.\ J.\ C {\bf 72} (2012) 2172
  [arXiv:1208.0934 [hep-ph]].

\bibitem{WMAP}
  G.~Hinshaw {\it et al.}  [WMAP Collaboration],
  arXiv:1212.5226 [astro-ph.CO].

\bibitem{Planck}
  P.~A.~R.~Ade {\it et al.}  [Planck Collaboration],
  arXiv:1303.5076 [astro-ph.CO].

\bibitem{bsgSM_NLO}
  A.~J.~Buras, A.~Czarnecki, M.~Misiak and J.~Urban,
  Nucl.\ Phys.\ B {\bf 631} (2002) 219
  [hep-ph/0203135].

\bibitem{Bsmumu_CMS}
  S.~Chatrchyan {\it et al.}  [CMS Collaboration],
  Phys.\ Rev.\ Lett.\  {\bf 111} (2013) 101804
  [arXiv:1307.5025 [hep-ex]].

\bibitem{Bsmumu_LHCb}
  RAaij {\it et al.}  [LHCb Collaboration],
  Phys.\ Rev.\ Lett.\  {\bf 111} (2013) 101805
  [arXiv:1307.5024 [hep-ex]].

\bibitem{CMS_MA}
CMS Collaboration, CMS-PAS-HIG-12-050.

\bibitem{Allanach_higgs}
  B.~C.~Allanach, A.~Djouadi, J.~L.~Kneur, W.~Porod and P.~Slavich,
  JHEP {\bf 0409} (2004) 044
  [arXiv:hep-ph/0406166].


\bibitem{Higgs3loop}
  R.~V.~Harlander, P.~Kant, L.~Mihaila and M.~Steinhauser,
  Phys.\ Rev.\ Lett.\  {\bf 100} (2008) 191602
   [Phys.\ Rev.\ Lett.\  {\bf 101} (2008) 039901]
  [arXiv:0803.0672 [hep-ph]];
  P.~Kant, R.~V.~Harlander, L.~Mihaila and M.~Steinhauser,
  JHEP {\bf 1008} (2010) 104
  [arXiv:1005.5709 [hep-ph]];
  J.~L.~Feng, P.~Kant, S.~Profumo and D.~Sanford,
  arXiv:1306.2318 [hep-ph].

\bibitem{YUmirage}
  A.~Anandakrishnan and S.~Raby,
  arXiv:1303.5125 [hep-ph].

\bibitem{Martin}
  S.~P.~Martin,
  Phys.\ Rev.\  {\bf D79 } (2009)  095019.
  [arXiv:0903.3568 [hep-ph]].

\bibitem{Moroi}
T.~Moroi,
  Phys.\ Rev.\  D {\bf 53} (1996) 6565
  [Erratum-ibid.\  D {\bf 56} (1997) 4424]
  [arXiv:hep-ph/9512396].

\bibitem{Stockinger}
  D.~Stockinger,
  J.\ Phys.\ G {\bf G34 } (2007)  R45-R92.
  [hep-ph/0609168].

\bibitem{F1F24King}
  S.~F.~King, J.~P.~Roberts and D.~P.~Roy,
  JHEP {\bf 0710} (2007) 106
  [arXiv:0705.4219 [hep-ph]].

\bibitem{F1F24Martin}
  J.~E.~Younkin and S.~P.~Martin,
  Phys.\ Rev.\ D {\bf 85} (2012) 055028
  [arXiv:1201.2989 [hep-ph]];
  S.~P.~Martin,
  Phys.\ Rev.\ D {\bf 75} (2007) 115005
  [hep-ph/0703097 [HEP-PH]].

\bibitem{Slansky}
  R.~Slansky,
  Phys.\ Rept.\  {\bf 79} (1981) 1.

\bibitem{BNL}
  G.~W.~Bennett {\it et al.} [ Muon G-2 Collaboration ],
  Phys.\ Rev.\  {\bf D73 } (2006)  072003.
  [hep-ex/0602035].

\bibitem{Davier_g2SM}
  M.~Davier, A.~Hoecker, B.~Malaescu, Z.~Zhang,
  Eur.\ Phys.\ J.\  {\bf C71 } (2011)  1515.
  [arXiv:1010.4180 [hep-ph]].

\bibitem{Higgs_IMH}
  M.~Badziak, E.~Dudas, M.~Olechowski and S.~Pokorski,
  JHEP {\bf 1207} (2012) 155
  [arXiv:1205.1675 [hep-ph]].

\bibitem{welltempered}
  N.~Arkani-Hamed, A.~Delgado, G.~F.~Giudice,
  Nucl.\ Phys.\  {\bf B741 } (2006)  108-130.
  [hep-ph/0601041].

\bibitem{winoLSP_charginosearch}
  T.~Gherghetta, G.~F.~Giudice and J.~D.~Wells,
  Nucl.\ Phys.\ B {\bf 559} (1999) 27
  [hep-ph/9904378].

\bibitem{LHCreach}
  H.~Baer, V.~Barger, A.~Lessa and X.~Tata,
  Phys.\ Rev.\ D {\bf 86} (2012) 117701
  [arXiv:1207.4846 [hep-ph]].

\bibitem{Chankowski}
  P.~H.~Chankowski, J.~R.~Ellis, M.~Olechowski, S.~Pokorski,
  Nucl.\ Phys.\  {\bf B544 } (1999)  39-63.
  [hep-ph/9808275].


\bibitem{Okumura}
  K.~-i.~Okumura, L.~Roszkowski,
  Phys.\ Rev.\ Lett.\  {\bf 92 } (2004)  161801.
  [hep-ph/0208101].

\bibitem{YU_Hall}
  G.~Elor, L.~J.~Hall, D.~Pinner and J.~T.~Ruderman,
  JHEP {\bf 1210} (2012) 111
  [arXiv:1206.5301 [hep-ph]].

\bibitem{ATLAS_wino_tau}
ATLAS Collaboration, ATLAS-CONF-2013-028.

\bibitem{CMS_wino}
CMS Collaboration, CMS-PAS-SUS-12-022.

\bibitem{Atlas_sbottom}
  ATLAS Collaboration,
  ATLAS-CONF-2013-053.

\bibitem{mA_prospects}
  A.~Arbey, M.~Battaglia and F.~Mahmoudi,
  Phys.\ Rev.\ D {\bf 88} (2013) 015007
  [arXiv:1303.7450 [hep-ph]].

\bibitem{Susyhit}
  A.~Djouadi, M.~M.~Muhlleitner and M.~Spira,
  Acta Phys.\ Polon.\ B\ {\bf 38} (2007) 635
  [hep-ph/0609292];
  A.~Djouadi, J.~Kalinowski and M.~Spira,
  Comput.\ Phys.\ Commun.\ \ {\bf 108} (1998) 56
  [hep-ph/9704448];
  M.~Muhlleitner, A.~Djouadi and Y.~Mambrini,
  Comput.\ Phys.\ Commun.\ \ {\bf 168} (2005) 46
  [hep-ph/0311167].

\bibitem{LHC_lumi}
ATLAS Collaboration, ATL-PHYS-PUB-2012-004



\bibitem{NMSSMmixing}
  M.~Badziak, M.~Olechowski and S.~Pokorski,
  JHEP {\bf 1306} (2013) 043
  [arXiv:1304.5437 [hep-ph]].



























\end{thebibliography}
\end{document}